\documentclass[10pt]{article}

\usepackage[dvips]{graphicx}

\usepackage{color}
\definecolor{oneblue}{rgb}{0,0.0,0.75}

\usepackage{xspace} 
\usepackage{amsmath,amssymb}

\usepackage{fancyhdr}
\def\Dv{\mathbf{D}}
\def\kv{\mathbf{k}}
\def\uv{\mathbf{u}}
\def\wv{\mathbf{w}}
\def\Fv{\mathbf{F}}
\def\Av{\mathbf{A}}
\def\fv{\mathbf{f}}

\def\Gv{\mathbf{G}}
\def\Sv{\mathbf{S}}
\def\qv{\mathbf{q}}
\def\R{\mathbb{R}}
\def\Z{\mathbb{Z}}
\def\N{\mathbb{N}}

\newcommand{\pd}[2]{\frac{\partial#1}{\partial#2}}
\newcommand{\od}[2]{\frac{d#1}{d#2}}
\newcommand{\god}[2]{\frac{D#1}{D#2}}
\newcommand{\tens}[1]{\mathbf{#1}}

\newcommand{\abs}[1]{\left|#1\right|}

\title{Comparison between three-dimensional linear and nonlinear tsunami generation models}
\author{Youen Kervella \footnote{IFREMER, Laboratoire DYNECO/PHYSED, BP 70, 29280 Plouzan\'e, France} \and
Denys Dutykh \footnote{Centre de Math\'ematiques et de Leurs Applications,
\'{E}cole Normale Sup\'erieure de Cachan, 61 avenue du Pr\'esident Wilson, 94235 Cachan cedex, France;
e-mail: dias@cmla.ens-cachan.fr, phone: +33 1 47 40 59 00, fax: +33 1 47 40 59 01} 
\and Fr\'ed\'eric Dias \footnotemark[2]}
\date{1 February 2007}

\begin{document}
\maketitle

\begin{abstract}
The modeling of tsunami generation is an essential phase in understanding tsunamis. For tsunamis generated by underwater
earthquakes, it involves the modeling of the sea bottom motion as well as the resulting motion of the water above it. A comparison 
between various models for three-dimensional water motion,
ranging from linear theory to fully nonlinear theory, is performed. It is found that for most events the linear
theory is sufficient. However, in some cases, more sophisticated theories are needed. Moreover, it is shown that
the passive approach in which the seafloor deformation is simply translated to the ocean surface is not always equivalent
to the active approach in which the bottom motion is taken into account, even if the deformation is supposed to be instantaneous. 
\end{abstract}


\section {Introduction}

Tsunami wave modeling is a challenging task. In particular, it is essential to understand the first minutes of a tsunami, its
propagation and finally the resulting inundation and impact on structures. The focus of the present paper is on the generation process. 
There are different natural phenomena that can lead to a tsunami. For example, one can mention submarine mass failures, slides, 
volcanic eruptions, falls of asteroids, etc. We refer to the recent review on tsunami science \cite{Syno2006} for a 
complete bibliography on the topic. The present work focuses on tsunami generation by earthquakes. 

Two steps in modeling are necessary for an accurate description of tsunami generation: a model for the earthquake 
fed by the various seismic parameters, and a model for the formation of surface gravity waves resulting from the deformation 
of the seafloor. In the absence of sophisticated source models, one often uses analytical solutions based on dislocation theory
in an elastic half-space for the seafloor displacement \cite{Okada85}. For the resulting water motion, the standard practice 
is to transfer the inferred seafloor
displacement to the free surface of the ocean. In this paper, we will call this approach the {\it passive generation} approach.
\footnote{In the pioneering paper \cite{kajiura}, Kajiura analyzed the applicability of the passive approach using Green's functions.
In the tsunami literature, this approach is sometimes called the piston model of tsunami generation.}
This approach leads to a well-posed initial value problem with zero velocity. An open question for tsunami forecasting modelers is 
the validity of neglecting the initial velocity. In a recent note, Dutykh et al. \cite{ddk}
used linear theory to show that indeed differences may exist between the standard passive generation and the {\it active generation} 
that takes into account the dynamics of seafloor displacement. The transient wave generation due to the coupling between the seafloor 
motion and the free surface has been considered by a few authors only. One of the reasons is that it is commonly
assumed that the source details are not important.\footnote{As pointed out by Geist et al. \cite{Geist2006}, the 2004 Indian Ocean
tsunami shed some doubts about this belief. The measurements from land based stations that use the Global Positioning System 
to track ground movements revealed that the fault continued to slip after it stopped releasing seismic energy. Even though this
slip was relatively slow, it contributed to the tsunami and may explain the surprising tsunami heights.}
Ben-Menahem and Rosenman \cite{Ben-M} calculated the two-dimensional radiation pattern from a moving source
(linear theory). Tuck and Hwang \cite{Tuck} solved the linear long-wave equation in the presence of a moving bottom and a uniformly
sloping beach. Hammack \cite{Hammack} generated waves experimentally by raising or lowering a box at one end of a channel.  
According to Synolakis and Bernard \cite{Syno2006}, Houston and Garcia \cite{Houston} were the first to use more geophysically 
realistic initial conditions. For obvious
reasons, the quantitative differences in the distribution of seafloor displacement due to underwater earthquakes compared
with more conventional earthquakes are still poorly known. Villeneuve and Savage \cite{Savage} derived model equations which combine 
the linear effect of frequency dispersion and the nonlinear effect of amplitude dispersion, and included the effects of a moving bed. 
Todorovska and Trifunac \cite{todo} considered the generation of tsunamis by a slowly spreading uplift of the seafloor. 

In this paper, we mostly follow the standard passive generation approach. Several tsunami generation models and numerical methods 
suited for these models are presented and compared. The focus of our work is on modelling the fluid motion. It is assumed that the 
seabed deformation satisfies all the necessary hypotheses required to apply Okada's solution.
The main objective is to confirm or infirm the lack of importance of nonlinear effects and/or frequency dispersion in tsunami generation. 
This result may have implications in terms of computational cost. The goal is to optimize the ratio between the complexity of the model 
and the accuracy of the results. Government agencies need to compute accurately tsunami propagation in real time in order to know where
to evacuate people. Therefore any saving in computational time is crucial (see for example the code MOST used by the National Oceanic
and Atmospheric Administration in the US \cite{TS} or the code TUNAMI developed by the Disaster Control Research Center in Japan). 
Liu and Liggett \cite{Liu} already performed comparisons between linear and nonlinear water waves but their study was restricted to simple bottom 
deformations, namely the generation of transient waves by an upthrust of a rectangular block, and the nonlinear computations 
were restricted to two-dimensional flows. Bona et al. \cite{BPS} assessed how well a model equation with weak nonlinearity
and dispersion describes the propagation of surface water waves generated at one end of a long channel. In their
experiments, they found that the inclusion of a dissipative term was more important than the inclusion of nonlinearity,
although the inclusion of nonlinearity was undoubtedly beneficial in describing the observations. The importance of
dispersive effects in tsunami propagation is not directly addressed in the present paper. Indeed these effects cannot be measured 
without taking into account the duration (or distance) of tsunami propagation \cite{Ernie}.

The paper is organized as follows. In Section 2, we review the equations that are commonly used for water-wave propagation, 
namely the fully nonlinear potential flow (FNPF) equations. Section 3 provides a description of the linear theory, with explicit
expressions for the free-surface elevation and the velocities everywhere inside the fluid domain, both for active and passive
generations. Section 4 is devoted to the nonlinear shallow water (NSW) equations and their numerical integration by a finite
volume scheme. In Section 5 we briefly
describe the boundary element numerical method used to integrate the FNPF equations. The following section (Section 6)
is devoted to comparisons between the various models and a discussion on the results. The main conclusion is that linear
theory is sufficient in general but that passive generation overestimates the initial transient waves in some cases. Finally 
directions for future research are outlined.

\section {Physical problem description}

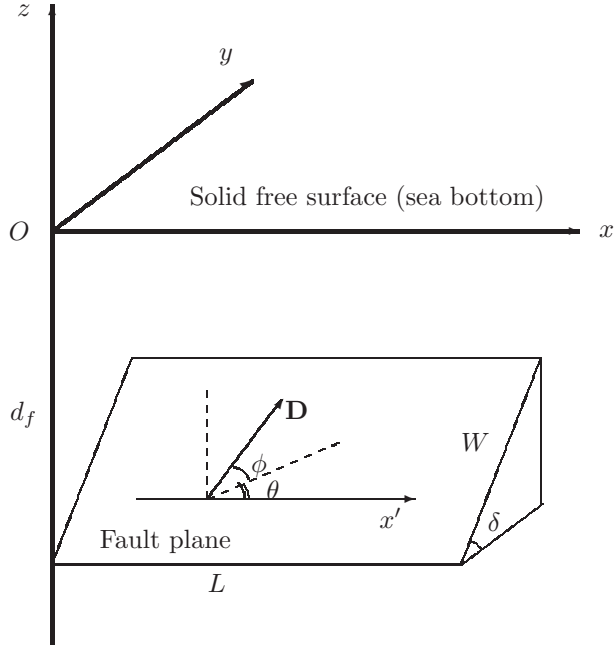
\begin{figure}[htbp]
\begin{center}
\unitlength 1mm
\def\Dv{\mathbf{D}}

\begin{picture}(83.75,84.38)(0,0)

\linethickness{0.55mm}
\put(10.00,54.38){\line(1,0){70.00}}
\put(80.00,54.38){\vector(1,0){0.12}}

\linethickness{0.55mm}
\multiput(10.00,54.38)(0.16,0.12){167}{\line(1,0){0.16}}
\put(36.25,74.38){\vector(4,3){0.12}}

\put(83.75,54.38){\makebox(0,0)[cc]{$x$}}

\put(33.13,77.50){\makebox(0,0)[cc]{$y$}}

\linethickness{0.45mm}
\put(10.00,-0.63){\line(0,1){85.00}}
\put(10.00,84.38){\vector(0,1){0.12}}

\put(6.25,83.75){\makebox(0,0)[cc]{$z$}}
\put(6.25,30){\makebox(0,0)[cc]{$d_f$}}

\put(5.63,54.38){\makebox(0,0)[cc]{$O$}}

\linethickness{0.15mm}
\multiput(10.00,10.00)(0.12,0.31){89}{\line(0,1){0.31}}

\linethickness{0.15mm}
\put(10.00,10.00){\line(1,0){54.38}}

\linethickness{0.15mm}
\multiput(64.38,10.00)(0.12,0.31){89}{\line(0,1){0.31}}

\linethickness{0.15mm}
\multiput(64.38,10.00)(0.16,0.12){68}{\line(1,0){0.16}}

\linethickness{0.15mm}
\put(75.01,18.13){\line(0,1){19.37}}

\put(68.76,15.63){\makebox(0,0)[cc]{$\delta$}}

\linethickness{0.15mm}
\multiput(66.37,12.73)(0.12,-0.12){6}{\line(1,0){0.12}}
\multiput(65.42,13.10)(0.32,-0.12){3}{\line(1,0){0.32}}

\linethickness{0.25mm}
\multiput(30.63,18.75)(0.12,0.16){83}{\line(0,1){0.16}}
\put(40.63,31.88){\vector(3,4){0.12}}

\linethickness{0.15mm}
\multiput(30.63,18.75)(1.84,0.79){10}{\multiput(0,0)(0.31,0.13){3}{\line(1,0){0.31}}}

\put(25.01,13.13){\makebox(0,0)[cc]{Fault plane}}

\put(51.88,58.76){\makebox(0,0)[cc]{Solid free surface (sea bottom)}}

\put(42.51,30.63){\makebox(0,0)[cc]{$\Dv$}}

\linethickness{0.15mm}
\put(21.26,18.75){\line(1,0){36.87}}
\put(58.13,18.75){\vector(1,0){0.12}}

\put(55.01,16.25){\makebox(0,0)[cc]{$x'$}}

\linethickness{0.15mm}
\multiput(35.93,21.84)(0.13,-0.25){2}{\line(0,-1){0.25}}
\multiput(35.47,22.29)(0.12,-0.11){4}{\line(1,0){0.12}}
\multiput(34.84,22.66)(0.21,-0.12){3}{\line(1,0){0.21}}
\multiput(34.08,22.93)(0.38,-0.14){2}{\line(1,0){0.38}}

\put(37.51,23.13){\makebox(0,0)[cc]{$\phi$}}

\linethickness{0.15mm}
\multiput(35.55,19.64)(0.08,-0.68){1}{\line(0,-1){0.68}}
\multiput(35.26,20.29)(0.15,-0.32){2}{\line(0,-1){0.32}}
\multiput(34.76,20.86)(0.12,-0.14){4}{\line(0,-1){0.14}}

\linethickness{0.15mm}
\multiput(36.00,19.81)(0.12,-0.57){2}{\line(0,-1){0.57}}
\multiput(35.53,20.78)(0.12,-0.24){4}{\line(0,-1){0.24}}

\put(39.38,20.00){\makebox(0,0)[cc]{$\theta$}}

\put(49.38,14.38){\makebox(0,0)[cc]{}}

\linethickness{0.15mm}
\multiput(30.63,18.75)(0,1.92){8}{\line(0,1){0.96}}

\put(31.88,7.50){\makebox(0,0)[cc]{$L$}}

\put(66.26,26.25){\makebox(0,0)[cc]{$W$}}

\linethickness{0.15mm}
\put(20.63,37.50){\line(1,0){54.38}}

\end{picture}
\end{center}
  \caption{Geometry of the source model (dip angle $\delta$, depth $d_f$, length $L$, width $W$) and orientation of 
Burger's vector $\Dv$ (rake angle $\theta$, angle $\phi$ between the fault plane and Burger's vector).}\label{fig:1}
\end{figure}

\begin{figure}
\centerline{\includegraphics[width=1.2\linewidth]{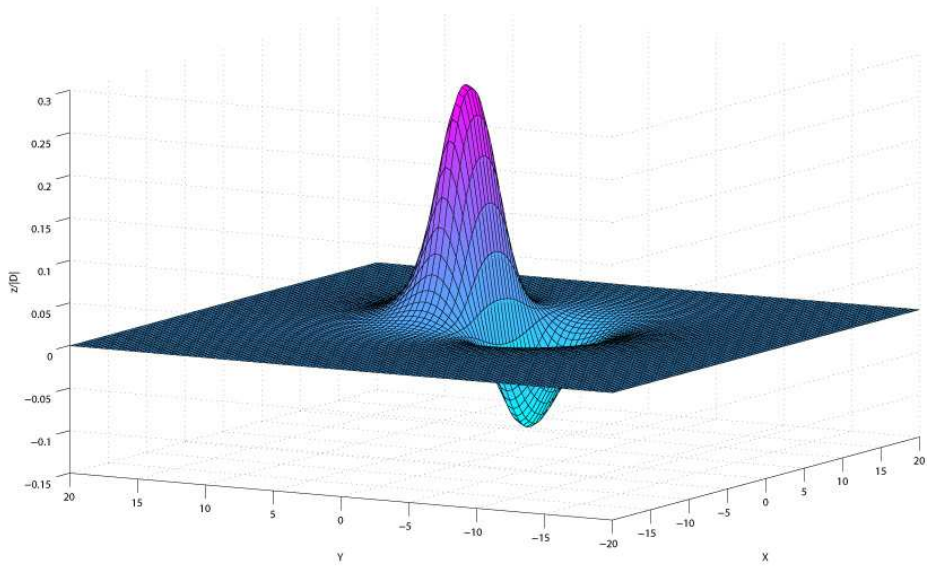}}
  \caption{Typical seafloor deformation due to dip-slip faulting. The parameters are those of Table \ref{parset}.
The distances along the horizontal axes $x$ and $y$ are expressed in kilometers.}
  \label{dip}
\end{figure}

In the whole paper, the vertical coordinate is denoted by $z$, while the two horizontal coordinates are denoted
by $x$ and $y$, respectively. 
The sea bottom deformation following an underwater earthquake is a complex phenomenon. This is why, for theoretical
or experimental studies, researchers have often used simplified bottom motions such as the vertical motion of a box. 
In order to determine the deformations of the sea bottom due to an earthquake,   
we use the analytical solution obtained for a dislocation in an elastic half-space \cite{Okada85}. This
solution, which at present time is used by the majority of tsunami wave modelers to produce an initial condition
for tsunami propagation simulations, provides an explicit expression of the bottom surface deformation that depends on a dozen of source 
parameters such as the dip angle $\delta$, fault depth $d_f$, fault dimensions (length and width), Burger's vector $\Dv$, Young's
modulus, Poisson ratio, etc. 
Some of these parameters are shown in figure \ref{fig:1}. More details can be found in \cite{Dias2} for example.
A value of $90^\circ$ for the dip angle corresponds to a vertical fault. Varying the fault slip $|\Dv|$ does not change
the co-seismic deformation pattern, only its magnitude. The values of the parameters used in the present paper are given 
in Table \ref{parset}. A typical dip-slip
solution is shown in figure \ref{dip} (the angle $\phi$ is equal to 0, while the rake angle $\theta$ is equal to $\pi/2$).
\begin{table} 
\begin{center}
\begin{tabular}{lc}
{\it parameter} & {\it value} \\
\hline
  Dip angle $\delta$ & $13^\circ$ \\
  Fault depth $d_f$, km & 3 \\
  Fault length $L$, km & 6 \\
  Fault width $W$, km & 4 \\
  Magnitude of Burger's vector $|\Dv|$, m & 1 \\
  Young's modulus $E$, GPa & 9.5 \\
  Poisson ratio $\nu$ & 0.23 \\
\hline
\end{tabular}
\end{center}
\caption[]{Typical parameter set for the source used to model the seafloor deformation due to an 
earthquake in the present study. The dip angle, Young's modulus and Poisson ratio correspond roughly to those of the 2004
Sumatra event. The fault depth, length and width, as well as the magnitude of Burger's vector, have been reduced for
computation purposes.}
\label{parset}
\end{table}

\begin{figure}
\centerline{\includegraphics[width=1.2\linewidth]{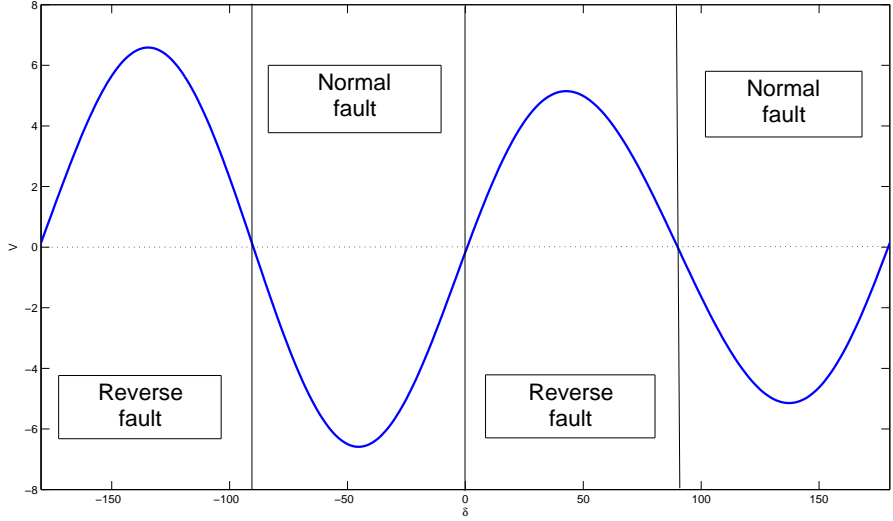}}
  \caption{Initial net volume $V$ (in km$^3$) of the seafloor displacement as a function of the dip angle $\delta$ (in $^\circ$).
All the other parameters, which are given in Table \ref{parset}, are kept constant.}
  \label{fig:dipangle}
\end{figure}

Let $z=\zeta(x,y,t)$ denote the deformation of the sea bottom. Hammack and Segur \cite {Hammack2} suggested that there are two main 
kinds of behaviour for the generated waves depending on whether the net volume $V$ of the initial bottom surface deformation 
	\begin{equation*}
  V =  \int\limits_{\mathbb{R}^2} \zeta (x,y,0)\,dx dy 
\end{equation*}
	is positive or not.\footnote{However it should be noted that the analysis of \cite{Hammack2} is restricted to 
one-dimensional uni-directional waves. We assume here that their conclusions can be extended to two-dimensional 
bi-directional waves.} 
A positive $V$ is achieved for example for a ``reverse fault'', i.e. when the dip angle $\delta$ satisfies 
$0 \leq \delta \leq \pi /2$ or $-\pi \leq \delta \leq -\pi /2 $, as shown in figure \ref{fig:dipangle}.
A negative $V$ is achieved for a ``normal fault'', i.e. when the dip angle $\delta$ satisfies $\pi/2 \leq \delta \leq \pi$ 
or $-\pi/2 \leq \delta \leq 0$.

\begin{figure}
\centerline{\includegraphics[width=1.2\linewidth]{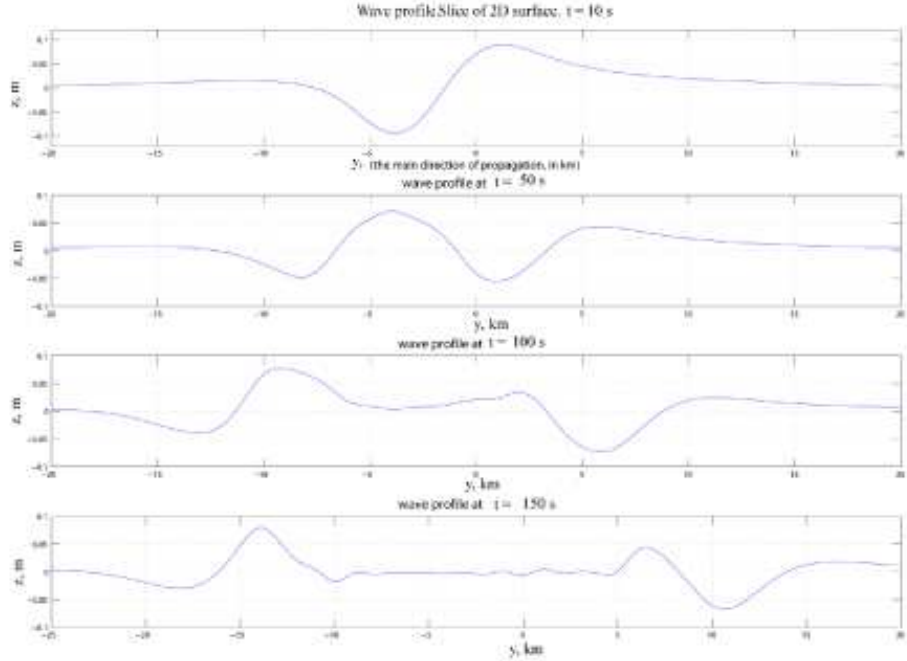}}
  \caption{Wave profiles at different times for the case of a normal fault ($\delta=167^\circ$). The seafloor
deformation occurs instantaneously at $t=0$. The water depth $h(x,y)$ is assumed to be constant.}
  \label{fig:normalf}
\end{figure}

The conclusions of \cite{Hammack2} are based on the Korteweg--de Vries (KdV) equation and were in part confirmed by their
experiments. If $V$ is positive, waves of stable form (solitons) evolve and are followed by a dispersive train of oscillatory waves,
regardless of the exact structure of $\zeta(x,y,0)$. 
If $V$ is negative, and if the initial data is non-positive everywhere, no solitons evolve.
But, if $V$ is negative and there is a region of elevation in the initial data (which corresponds to a typical Okada solution for
a normal fault), solitons can evolve and we have checked this last result using the FNPF equations (see figure \ref{fig:normalf}).
In this study, we focus on the case where $V$ is positive with a dip angle $\delta$ equal to $13^\circ$, according 
to the seismic data of the 26 December 2004 Sumatra-Andaman event (see for example \cite{Lay}). However, the sea bottom deformation 
often has an $N-$shape, with subsidence on one side of the fault and uplift on the other side as shown in figure \ref{dip}. In that 
case, one may expect the positive $V$ behaviour on one side and the negative $V$ behaviour on the other side. Recall that the 
experiments of Hammack and Segur \cite{Hammack2} were performed in the
presence of a vertical wall next to the moving bottom and their analysis was based on the uni-directional KdV wave equation. 

\begin{figure}
\centerline{\includegraphics[width=0.9\linewidth]{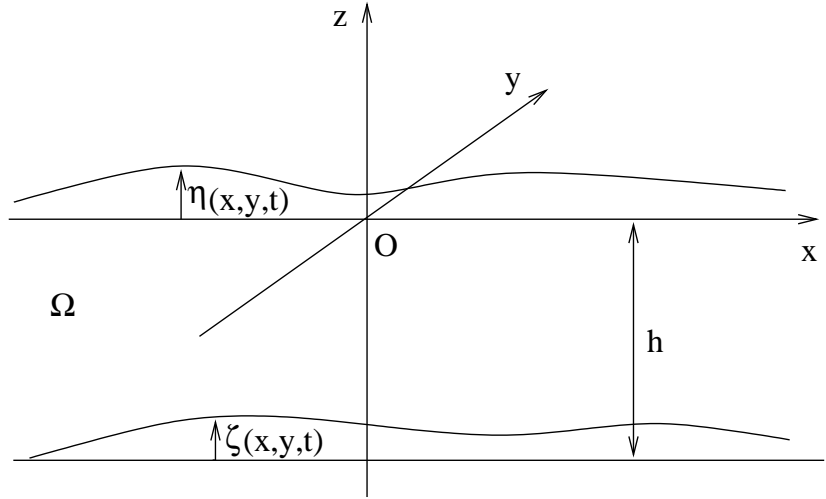}}
  \caption{Definition of the fluid domain $\Omega$ and of the coordinate system $(x,y,z)$.}
  \label{fig:fluid}
\end{figure}

We now consider the fluid domain. A sketch is shown in figure
\ref{fig:fluid}. The fluid domain $\Omega$ is bounded above by the free surface and below
by the rigid ocean floor. It is unbounded in the horizontal $x-$ and $y-$
directions. So, one can write
$$
  \Omega = \mathbb{R}^2\times[-h(x,y)+\zeta(x,y,t),\eta(x,y,t)].
$$
Before the earthquake the fluid is assumed to be at rest, thus the
free surface and the solid boundary are defined by $z=0$ and $z=-h(x,y)$,
respectively. For simplicity $h(x,y)$ is assumed to be a constant. Of course, in real situations, this is never the case
but for our purpose the bottom bathymetry is not important. 
Starting at time $t=0$, the solid boundary moves in a prescribed manner which is given by
\begin{equation*}
    z = -h + \zeta(x,y,t), \quad t \ge 0.
\end{equation*}

The deformation of the sea bottom is assumed to have all the necessary properties
needed to compute its Fourier transform in $x,y$ and its Laplace transform in $t$. 
The resulting deformation of the free surface
$z=\eta(x,y,t)$ is to be found as part of the solution. It is also assumed that the
fluid is incompressible and the flow irrotational. The latter
implies the existence of a velocity potential $\phi(x,y,z,t)$ which
completely describes the flow. By definition of $\phi$ the fluid
velocity vector can be expressed as $\qv = \nabla\phi$. Thus, the
continuity equation becomes
\begin{equation} \label{Laplace3}
  \nabla\cdot \qv = \Delta\phi = 0, \quad (x,y,z) \in \Omega.
\end{equation}
The potential $\phi(x,y,z,t)$ must satisfy the following
kinematic boundary conditions on the free surface and the solid
boundary, respectively:
\begin{eqnarray*}
  \pd{\phi}{z} &=& \pd{\eta}{t} + \pd{\phi}{x}\pd{\eta}{x} +
  \pd{\phi}{y}\pd{\eta}{y}, \quad z=\eta(x,y,t), \\
  \pd{\phi}{z} &=& \pd{\zeta}{t} + \pd{\phi}{x}\pd{\zeta}{x} +
  \pd{\phi}{y}\pd{\zeta}{y}, \quad z=-h + \zeta(x,y,t).
\end{eqnarray*}

Further assuming the flow to be inviscid and neglecting surface
tension effects, one can write the dynamic condition to be satisfied on the
free surface as
\begin{equation} \label{dc1}
  \pd{\phi}{t} + \frac12|\nabla\phi|^2 + g\eta = 0, \quad
  z=\eta(x,y ,t),
\end{equation}
where $g$ is the acceleration due to gravity. The atmospheric pressure has been chosen as reference pressure.

The equations are more transparent when written in dimensionless
variables. However the choice of the reference lengths and speeds is subtle. Different choices lead to different
models. Let the new independent variables be
\begin{equation*}
    \tilde{x} = x/\lambda, \quad \tilde{y} = y/\lambda, \quad \tilde{z} = z/d, \quad
    \tilde{t} = c_0 t/\lambda,
\end{equation*}
where $\lambda$ is the horizontal scale of the motion and $d$ a typical water depth. The speed $c_0$ is the long wave speed based
on the depth $d$ ($c_0 = \sqrt{gd}$). Let the new dependent variables be
\begin{equation*}
  \tilde{\eta} = \frac{\eta}a, \quad \tilde{\zeta} = \frac{\zeta}{a}, \quad
  \tilde{\phi} = \frac{c_0}{a g\lambda}\phi,
\end{equation*}
where $a$ is a characteristic wave amplitude.

In dimensionless form, and after dropping the tildes, the equations
become
\begin{equation}\label{lapladim}
  \pd{^2\phi}{z^2} + \mu^2 \left(\pd{^2\phi}{x^2} + 
  \pd{^2\phi}{y^2}\right)= 0, \quad (x,y,z) \in \Omega,
\end{equation}
\begin{equation}\label{kinfreesurfadim}
  \pd{\phi}{z} = \mu^2 \pd{\eta}{t} + \varepsilon \mu^2 \left(
  \pd{\phi}{x}\pd{\eta}{x} + \pd{\phi}{y}\pd{\eta}{y}
  \right), \quad z = \varepsilon \eta(x,y,t), 
\end{equation}
\begin{equation}\label{kinsolbadim}
  \pd{\phi}{z} = \mu^2 \pd{\zeta}{t} + \varepsilon \mu^2 \left(
  \pd{\phi}{x}\pd{\zeta}{x} + \pd{\phi}{y}\pd{\zeta}{y}
  \right), \quad z = -\frac{h}{d} + \varepsilon \zeta(x,y,t), 
\end{equation}
\begin{equation}\label{dynfreesurfadim}
	\mu^2 \pd{\phi}{t} + \frac12 \varepsilon  \left( \mu^2 \left(\pd{\phi}{x}\right)^2
	+\mu^2 \left(\pd{\phi}{y}\right)^2+\left(\pd{\phi}{z}\right)^2\right) + \mu^2
  \eta = 0, \quad z = \varepsilon \eta(x,y,t),
\end{equation}
where two dimensionless numbers have been introduced:
\begin{equation} \label{sdnumbers}
	\varepsilon = a/d , \quad \mu = d/\lambda.
 \end{equation}
For the propagation of tsunamis, both numbers $\varepsilon$ and $\mu$ are small. Indeed the satellite altimetry 
observations of the 2004 Boxing Day tsunami waves
obtained by two satellites that passed over the Indian Ocean a couple of hours after the
rupture process occurred gave an amplitude $a$ of roughly 60 cm in the open ocean. The typical 
wavelength estimated from the width of the 
segments that experienced slip is between 160 and 240 km \cite{Lay}. The water depth ranges from 4 km
towards the west of the rupture to 1 km towards the east. Therefore average values for $\varepsilon$ and $\mu$ in the open ocean 
are $\varepsilon \approx 2 \times 10^{-4}$ and $\mu \approx 2 \times 10^{-2}$.  A more precise range for these 
two dimensionless numbers is
\begin{equation}
1.5 \times 10^{-4} < \varepsilon < 6 \times 10^{-4}, \quad 4 \times 10^{-3} < \mu < 2.5 \times 10^{-2}.
\end{equation}

The water-wave problem, either in the form of an initial value problem (IVP) or in the form
of a boundary value problem (BVP), is difficult to solve because of the
nonlinearities in the boundary conditions and the unknown computational domain. 

\section {Linear theory}

First we perform the linearization of the above equations and boundary
conditions. It is equivalent to taking the limit of (\ref{lapladim})--(\ref{dynfreesurfadim}) as $\varepsilon \to 0$. 
The linearized problem can also be obtained by expanding the unknown functions as power series of the small parameter
$\varepsilon$. Collecting terms of the lowest order in
$\varepsilon$ yields the linear approximation. For the sake of convenience,
we now switch back to the physical variables. The linearized
problem in dimensional variables reads
\begin{equation}\label{lapl}
  \Delta \phi = 0, \qquad (x,y,z) \in \mathbb{R}^2\times[-h, 0],
\end{equation}
\begin{equation}\label{kinfreesurf}
  \pd{\phi}{z} = \pd{\eta}{t}, \qquad z = 0,
\end{equation}
\begin{equation}\label{kinsolb}
  \pd{\phi}{z} = \pd{\zeta}{t}, \qquad z = -h,
\end{equation}
\begin{equation}\label{dynfreesurf}
  \pd{\phi}{t} + g\eta = 0, \qquad z = 0.
\end{equation}
The bottom motion appears in equation (\ref{kinsolb}). 
Combining equations (\ref{kinfreesurf}) and (\ref{dynfreesurf})
yields the single free-surface condition
\begin{equation}\label{singlefreesurf}
  \pd{^2\phi}{t^2} + g\pd{\phi}{z} = 0,
  \qquad z = 0.
\end{equation}

Most studies of tsunami generation assume that the initial free-surface deformation 
is equal to the vertical displacement of the ocean bottom and take a zero velocity field as initial condition. 
The details of wave motion are completely neglected during the time that the source operates. While tsunami modelers often 
justify this assumption by the fact that the earthquake rupture
occurs very rapidly, there are some specific cases where the time scale and/or the horizontal extent of the bottom deformation 
may become an important factor. This was emphasized for example by Todorovska and Trifunac \cite{todo} and Todorovska et al. 
\cite{todo2}, who considered the generation of tsunamis by a slowly spreading uplift
of the seafloor in order to explain some observations related to past tsunamis. However they did not use realistic source models.

Our claim is that it is important to make a distinction between two mechanisms of generation: an active mechanism in which the bottom 
moves according to a given time law and a passive mechanism in which the seafloor deformation is simply translated to the free surface. 
Recently Dutykh et al. \cite{ddk} showed that even in the case
of an instantaneous seafloor deformation, there may be differences between these two generation processes.

\subsection{Active generation}

Since in this case the system is assumed to be at rest at $t=0$, the initial condition simply is
\begin{equation}\label{initialcond}
  \eta(x,y,0) \equiv 0.
\end{equation}
In fact, $\eta(x,y,t)=0$ for all times $t<0$ and the same condition holds for the velocities. For $t<0$,
the water is at rest and the bottom motion is such that $\zeta(x,y,t)=0$ for $t<0$.

The problem (\ref{lapl})--(\ref{singlefreesurf}) can be solved by using the method of integral transforms. We apply
the Fourier transform in $(x,y)$,
$$
  \mathfrak{F}[f] = \widehat{f}(k,\ell) = \int\limits_{\mathbb{R}^2} f(x,y)e^{-i(kx+\ell y)}\,dx dy, $$
  with its inverse transform
$$
  \mathfrak{F}^{-1}[\widehat{f}] = f(x,y) = \frac1{(2\pi)^2}
  \int\limits_{\mathbb{R}^2} \widehat{f}(k,\ell)
  e^{i(kx+\ell y)}\,dk d\ell,
$$
and the Laplace transform in time $t$,
$$
  \mathfrak{L}[g] = \tens{g}(s) = \int\limits_0^{+\infty}
  g(t) e^{-st}\, dt.
$$ 
For the combined Fourier and Laplace transforms, the following notation is introduced:
$$ \mathfrak{F}\mathfrak{L}[F(x,y,t)] = \overline{F}(k,\ell,s) = \int\limits_{\mathbb{R}^2}
e^{-i(kx+\ell y)}\,dx dy \int\limits_0^{+\infty} F(x,y,t)
  e^{-st}\, dt. $$ 
After applying the transforms, equations (\ref{lapl}), (\ref{kinsolb}) and
(\ref{singlefreesurf}) become
\begin{equation}\label{lapltrans}
  \od{^2\overline{\phi}}{z^2} - (k^2+\ell^2)\overline{\phi} = 0,
\end{equation}
\begin{equation}\label{kinsolbtrans}
  \od{\overline{\phi}}{z}(k,\ell,-h,s) = s\overline{\zeta}(k,\ell,s),
\end{equation}
\begin{equation}\label{freesurftrans}
  s^2\overline{\phi}(k,\ell,0,s) + g\od{\overline{\phi}}{z} (k,\ell,0,s) = 0.
\end{equation}
The transformed free-surface elevation can be obtained from (\ref{dynfreesurf}):
\begin{equation}\label{transfreesurf}
  \overline{\eta}(k,\ell,s) = -\frac{s}{g}\overline{\phi}(k,\ell,0,s).
\end{equation}
A general solution of equation (\ref{lapltrans}) is given by
\begin{equation}\label{genlapltrans}
  \overline{\phi}(k,\ell,z,s) = A(k,\ell,s)\cosh(mz) + B(k,\ell,s)\sinh(mz),
\end{equation}
where $m = \sqrt{k^2+\ell^2}$. The functions $A(k,\ell,s)$ and $B(k,\ell,s)$
can be easily found from the boundary conditions (\ref{kinsolbtrans}) and
(\ref{freesurftrans}):
\begin{eqnarray*}
  A(k,\ell,s) & = & -\frac{gs\overline{\zeta}(k,\ell,s)}{\cosh(mh)[s^2+gm\tanh(mh)]}, \\
  B(k,\ell,s) & = & \frac{s^3\overline{\zeta}(k,\ell,s)}{m\cosh(mh)[s^2+gm\tanh(mh)]}.
\end{eqnarray*}
From now on, the notation
\begin{equation}
\omega = \sqrt{gm\tanh(mh)}
\label{disprelation}
\end{equation}
will be used. Substituting the expressions for the functions $A$ and $B$ in (\ref{genlapltrans}) yields
\begin{equation}
  \overline{\phi}(k,\ell,z,s) = -\frac{gs\overline{\zeta}(k,\ell,s)}
  {\cosh(mh)(s^2+\omega^2)}
  \left(\cosh(mz) - \frac{s^2}{gm}\sinh(mz)\right). \label{phihat}
\end{equation}
The free-surface elevation (\ref{transfreesurf}) becomes
$$
  \overline{\eta}(k,\ell,s) = \frac{s^2\overline{\zeta}(k,\ell,s)}
  {\cosh(mh)(s^2+\omega^2)}.
$$

Inverting the Laplace and Fourier transforms provides the general
integral solution
\begin{equation}\label{genintsol}
  \eta(x,y,t) = \frac{1}{(2\pi)^2}\int\!\!\!\int\limits_{\!\!\!\!\!\R^2}
  \frac{e^{i(kx+\ell y)}}{\cosh(mh)}\frac1{2\pi i}
  \int\limits_{\mu-i\infty}^{\mu+i\infty}
  \frac{s^2\overline{\zeta}(k,\ell,s)}{s^2+\omega^2}e^{st}ds\; dk d\ell.
\end{equation}

In some applications it is important to know not only the
free-surface elevation but also the velocity field inside the fluid
domain. In the present study we consider seabed deformations with the structure
\begin{equation}\label{specrepr}
  \zeta(x,y,t) := \zeta_0(x,y)T(t).
\end{equation}
Mathematically we separate the time dependence from the spatial coordinates.
There are two main reasons for doing this. First of all we want to
be able to invert analytically the Laplace transform. The
second reason is more fundamental. In fact, dynamic
source models are not easily available. Okada's solution, which was briefly described in the previous section,
provides the static sea-bed deformation $\zeta_0(x,y)$. Hammack \cite{Hammack} considered two types of time histories:
an exponential and a half-sine bed movements. Dutykh and Dias \cite{Dias2} considered two additional time histories: a linear and
an instantaneous bed movements. We show below that taking an instantaneous seabed
deformation (in that case the function $T(t)$ is the Heaviside step function) is not equivalent to instantaneously 
transferring the seabed deformation to the ocean surface\footnote{In the framework of the linearized shallow water
equations, one can show that it is equivalent to take an instantaneous seabed deformation or to instantaneously
transfer the seabed deformation to the ocean surface \cite{Tuck}.}. 

In equation (\ref{phihat}), we obtained the Fourier--Laplace transform of the velocity potential
$\phi(x,y,z,t)$:
\begin{equation}\label{potential}
  \overline{\phi}(k,\ell,z,s) = -\frac{gs\widehat{\zeta_0}(k,\ell)\tens{T}(s)}{\cosh(m h)(s^2+\omega^2)}
  \left(\cosh(m z) - \frac{s^2}{gm}\sinh(m z)\right).
\end{equation}
Let us evaluate the velocity field at an arbitrary level $z=\beta h$
with $-1\leq\beta\leq 0$. In the linear approximation the
value $\beta = 0$ corresponds to the free surface while $\beta=-1$ corresponds to the
bottom. Below the horizontal
velocities are denoted by $\uv$ and the horizontal gradient $(\partial/\partial x,\partial/\partial y)$ is denoted by
$\nabla_h$. The vertical velocity component is simply $w$. The Fourier transform parameters are denoted by
$\kv = (k,\ell)$.

Taking the Fourier and Laplace transforms of 
$$
  \uv(x,y,t;\beta) = \left.\nabla_h \phi(x,y,z,t)\right|_{z=\beta h}
$$
yields
\begin{eqnarray*}
\overline{\uv}(k,\ell,s;\beta) & = & -i\overline{\phi}(k,\ell,\beta h, s)\kv \\
 & = & i\frac{gs\widehat{\zeta_0}(k,\ell)\tens{T}(s)}{\cosh(m h)(s^2+\omega^2)}
 \left(\cosh(\beta mh) - \frac{s^2}{gm}\sinh(\beta mh)
  \right)\kv.
\end{eqnarray*}
Inverting the Fourier and Laplace transforms gives the general formula for the
horizontal velocity vector:
\begin{eqnarray*}
  \uv(x,y,t;\beta) & = &
  \frac{ig}{4\pi^2}\int\!\!\!\int\limits_{\!\!\!\!\!\R^2}
  \frac{\kv \widehat{\zeta_0}(k,\ell)\cosh(m\beta h) e^{i(kx+\ell y)}}{\cosh(m h)}
  \frac{1}{2\pi i}\int\limits_{\mu-i\infty}^{\mu+i\infty}
  \frac{s\tens{T}(s)e^{st}}{s^2+\omega^2}\;ds\; d\kv \\
  & & -\frac{i}{4\pi^2}\int\!\!\!\int\limits_{\!\!\!\!\!\R^2}
  \frac{\kv \widehat{\zeta_0}(k,\ell)\sinh(m\beta h) e^{i(kx+\ell y)}}{m\cosh(m h)}
  \frac{1}{2\pi i}\int\limits_{\mu-i\infty}^{\mu+i\infty}
  \frac{s^3\tens{T}(s)e^{st}}{s^2+\omega^2}\;ds\; d\kv.
\end{eqnarray*}

Next we determine the vertical component of the velocity
$w(x,y,t;\beta)$. It is easy to obtain the Fourier--Laplace transform
$\overline{w}(k,\ell,s;\beta)$ by differentiating (\ref{potential}):
$$
  \overline{w}(k,\ell,s;\beta) = \left.\pd{\overline{\phi}}{z}\right|_{z=\beta h} =
  \frac{sg\widehat{\zeta_0}(k,\ell)\tens{T}(s)}
  {\cosh(m h)(s^2+\omega^2)}\left(
  \frac{s^2}{g}\cosh(\beta mh) - m\sinh(\beta mh) \right).
$$
Inverting this transform yields
\begin{eqnarray*}
  w(x,y,t;\beta) & = & \frac1{4\pi^2}
  \int\!\!\!\int\limits_{\!\!\!\!\!\R^2}
  \frac{\cosh(\beta mh)\widehat{\zeta_0}(k,\ell)}{\cosh(m h)}e^{i(kx+\ell y)}
  \frac1{2\pi i}\int\limits_{\mu-i\infty}^{\mu+i\infty}
  \frac{s^3 \tens{T}(s)e^{st}}{s^2+\omega^2}\;ds\;d\kv \\
  & & -\frac{g}{4\pi^2}
  \int\!\!\!\int\limits_{\!\!\!\!\!\R^2}
  \frac{m\sinh(\beta mh)\widehat{\zeta_0}(k,\ell)}{\cosh(m h)}e^{i(kx+\ell y)}
  \frac1{2\pi i}\int\limits_{\mu-i\infty}^{\mu+i\infty}
  \frac{s\tens{T}(s)e^{st}}{s^2+\omega^2}\;ds\;d\kv,
\end{eqnarray*}
for $-1 \leq \beta \leq 0$. 

In the case of an instantaneous seabed deformation, $T(t)=H(t)$, where $H(t)$ denotes the Heaviside step function.
The resulting expressions for $\eta$, $\uv$ and $w$ (on the free surface), which are valid for $t>0$, are
\begin{eqnarray} \label{active_eta}
  \eta(x,y,t) & = & \frac{1}{(2\pi)^2}\int\!\!\!\int\limits_{\!\!\!\!\!\R^2}
  \frac{\widehat{\zeta_0}(k,\ell){e^{i(kx+\ell y)}}}{\cosh(mh)}  \cos\omega t \;  dk d\ell, \\
  \label{active_u} \uv(x,y,t;0) & = &
  \frac{ig}{4\pi^2}\int\!\!\!\int\limits_{\!\!\!\!\!\R^2}
  \frac{\kv \widehat{\zeta_0}(k,\ell)e^{i(kx+\ell y)}}{\cosh(mh)} \frac{\sin\omega t}{\omega} d\kv, \\
  \label{active_w} w(x,y,t;0) & = & -\frac{1}{4\pi^2}
  \int\!\!\!\int\limits_{\!\!\!\!\!\R^2}
  \frac{\widehat{\zeta_0}(k,\ell) e^{i(kx+\ell y)}}{\cosh(mh)} \omega \sin\omega t \; d\kv.
\end{eqnarray}
At time $t=0$, there is a singularity that can be incorporated in the above expressions. For simplicity, we only
consider the expressions for $t>0$.

Since tsunameters have one component that measures the pressure at the bottom (see for example \cite{Gonz}), it is interesting to 
provide as well the expression $p_b(x,y,t)$ for the pressure at the bottom. The pressure $p(x,y,z,t)$ can be obtained
from Bernoulli's equation, which was written explicitly for the free surface in 
equation (\ref{dc1}), but is valid everywhere in the fluid:
\begin{equation}
  \pd{\phi}{t} + \frac12|\nabla\phi|^2 + gz + \frac{p}{\rho} = 0. 
\label{bernou}
\end{equation}
After linearization, equation (\ref{bernou}) becomes
\begin{equation}
  \pd{\phi}{t} + gz + \frac{p}{\rho} = 0. 
\label{bernou_l}
\end{equation}
Along the bottom, it reduces to
\begin{equation}
  \pd{\phi}{t} + g(-h+\zeta) + \frac{p_b}{\rho} = 0, \qquad z=-h. 
\label{bernou_l_b}
\end{equation}
The time-derivative of the velocity potential is readily available in Fourier space. 
Inverting the Fourier and Laplace transforms and evaluating the resulting expression 
at $z=-h$ gives for an instantaneous seabed deformation
\begin{eqnarray*}
 \left.\pd{\phi}{t}\right|_{z=-h} & = & -\frac{g}{(2\pi)^2}\int\!\!\!\int\limits_{\!\!\!\!\!\R^2}
 \frac{\widehat{\zeta_0}(k,\ell)e^{i(kx+\ell y)}}{\cosh^2(mh)}\cos \omega t \; d\kv.  
\end{eqnarray*}
The bottom pressure deviation from the hydrostatic pressure is then given by
$$ p_b(x,y,t) = -\left.\rho\pd{\phi}{t}\right|_{z=-h} - \rho g\zeta. $$
Away from the deformed seabed, $\zeta$ goes to 0 so that $p_b$ simply is $-\left.\rho\phi_t\right|_{z=-h}$. The only
difference between $p_b$ and $\rho g\eta$ is the presence of an additional $\cosh(mh)$ term in the denominator of $p_b$.

\subsection{Passive generation}

In this case equation (\ref{kinsolb}) becomes
\begin{equation}\label{kinsolbbis}
  \pd{\phi}{z} = 0, \qquad z = -h,
\end{equation}
and the initial condition for $\eta$ now reads
\begin{equation*}
\eta(x,y,0) = \zeta_0(x,y),
\end{equation*}
where $\zeta_0(x,y)$ is the seafloor deformation. Initial velocities are assumed to be zero.

Again we apply the Fourier transform in the horizontal coordinates $(x,y)$. The Laplace transform is not applied since
there is no substantial dynamics in the problem. Equations (\ref{lapl}), (\ref{kinsolbbis}) and (\ref{singlefreesurf}) become
\begin{equation}\label{lapltransbis}
  \od{^2\widehat{\phi}}{z^2} - (k^2+\ell^2)\widehat{\phi} = 0,
\end{equation}
\begin{equation}\label{kinsolbtransbis}
  \od{\widehat{\phi}}{z}(k,\ell,-h,t) = 0,
\end{equation}
\begin{equation}\label{freesurftransbis}
  \pd{^2\widehat{\phi}}{t^2}(k,\ell,0,t) + g\pd{\widehat{\phi}}{z} (k,\ell,0,t) = 0.
\end{equation}
A general solution to Laplace's equation (\ref{lapltransbis}) is again given by
\begin{equation}\label{genlapltransbis}
  \widehat{\phi}(k,\ell,z,t) = A(k,\ell,t)\cosh(mz) + B(k,\ell,t)\sinh(mz),
\end{equation}
where $m = \sqrt{k^2+\ell^2}$.
The relationship between the functions $A(k,\ell,t)$ and $B(k,\ell,t)$
can be easily found from the boundary condition (\ref{kinsolbtransbis}):
\begin{equation}\label{relationbis}
  B(k,\ell,t) = A(k,\ell,t)\tanh(mh).
\end{equation}
From equation (\ref{freesurftransbis}) and the initial conditions one finds
\begin{equation}\label{Abis}
  A(k,\ell,t) = - \frac{g}{\omega} \widehat{\zeta_0}(k,\ell) \sin\omega t.
\end{equation}
Substituting the expressions for the functions $A$ and $B$ in (\ref{genlapltransbis}) yields
\begin{equation}\label{phihatbis}
  \widehat{\phi}(k,\ell,z,t) = -\frac{g}{\omega} \widehat{\zeta_0}(k,\ell)
  \sin\omega t \Bigl(\cosh(mz) + \tanh(mh) \sinh(mz)\Bigr). 
\end{equation}
From (\ref{dynfreesurf}), the free-surface elevation becomes
$$
  \widehat{\eta}(k,\ell,t) = \widehat{\zeta_0}(k,\ell) \cos\omega t.
$$
Inverting the Fourier transform provides the general integral solution
\begin{equation}\label{genintsolbis}
  \eta(x,y,t) = \frac{1}{(2\pi)^2}\int\!\!\!\int\limits_{\!\!\!\!\!\R^2}
  \widehat{\zeta_0}(k,\ell)  \cos\omega t \; {e^{i(kx+\ell y)}} dk d\ell.
\end{equation}

Let us now evaluate the velocity field in the fluid domain.
Equation (\ref{phihatbis}) gives the Fourier transform of the velocity potential
$\phi(x,y,z,t)$. Taking the Fourier transform of 
$$
  \uv(x,y,t;\beta) = \left.\nabla_h \phi(x,y,z,t)\right|_{z=\beta h}
$$
yields
\begin{eqnarray*}
\widehat{\uv}(k,\ell,t;\beta) & = & -i\widehat{\phi}(k,\ell,\beta h, t)\kv \\
 & = & i\frac{g}{\omega}\widehat{\zeta_0}(k,\ell)\sin\omega t  \Bigl(\cosh(\beta mh) + \tanh(mh) \sinh(\beta mh)\Bigr)\kv.
\end{eqnarray*}
Inverting the Fourier transform gives the general formula for the horizontal velocities
\begin{equation*}
  \uv(x,y,t;\beta) =
  \frac{ig}{4\pi^2}\int\!\!\!\int\limits_{\!\!\!\!\!\R^2}
  \kv \widehat{\zeta_0}(k,\ell)\frac{\sin\omega t}{\omega}\Bigl(\cosh(\beta mh) + \tanh(mh) \sinh(\beta mh)\Bigr) 
e^{i(kx+\ell y)} d\kv.
\end{equation*}
Along the free surface $\beta=0$, the horizontal velocity vector becomes
\begin{equation} \label{horvel}
  \uv(x,y,t;0) =
  \frac{ig}{4\pi^2}\int\!\!\!\int\limits_{\!\!\!\!\!\R^2}
  \kv \widehat{\zeta_0}(k,\ell)\frac{\sin\omega t}{\omega} e^{i(kx+\ell y)} d\kv.
\end{equation}

Next we determine the vertical component of the velocity
$w(x,y,t;\beta)$ at a given vertical
level $z=\beta h$. It is easy to obtain the Fourier transform
$\widehat{w}(k,\ell,t;\beta)$ by differentiating (\ref{phihatbis}):
$$
  \widehat{w}(k,\ell,t;\beta) = \left.\pd{\widehat{\phi}}{z}\right|_{z=\beta h} =
  - mg \frac{\sin\omega t}{\omega}\widehat{\zeta_0}(k,\ell) \Bigl(\sinh(\beta mh) + \tanh(mh) \cosh(\beta mh)\Bigr).
$$
Inverting this transform yields
\begin{multline*}
  w(x,y,t;\beta)  =  -\frac{g}{4\pi^2}
  \int\!\!\!\int\limits_{\!\!\!\!\!\R^2}
  \frac{m\sin\omega t}{\omega}\widehat{\zeta_0}(k,\ell) \Bigl(\sinh(\beta mh) + \\
  \tanh(mh) \cosh(\beta mh)\Bigr)e^{i(kx+\ell y)} d\kv
\end{multline*}
for $-1 \leq \beta \leq 0$. Using the dispersion relation, one can write
the vertical component of the velocity along the free surface ($\beta=0$) as
\begin{equation} \label{vervel}
  w(x,y,t;0) = -\frac{1}{4\pi^2}
  \int\!\!\!\int\limits_{\!\!\!\!\!\R^2}
  \omega \sin\omega t \; \widehat{\zeta_0}(k,\ell) e^{i(kx+\ell y)}
  d\kv.
\end{equation}
All the formulas obtained in this section are valid only if the integrals converge.

Again, one can compute the bottom pressure. At $z=-h$, one has
\begin{equation*}
 \left.\pd{\phi}{t}\right|_{z=-h} = -\frac{g}{(2\pi)^2}\int\!\!\!\int\limits_{\!\!\!\!\!\R^2}
 \frac{\widehat{\zeta_0}(k,\ell)e^{i(kx+\ell y)}}{\cosh(mh)}\cos \omega t \; d\kv.  
\end{equation*}
The bottom pressure deviation from the hydrostatic pressure is then given by
$$ p_b(x,y,t) = -\left.\rho\pd{\phi}{t}\right|_{z=-h} - \rho g\zeta. $$
Again, away from the deformed seabed, $p_b$ reduces to $-\left.\rho\phi_t\right|_{z=-h}$. The only
difference between $p_b$ and $\rho g\eta$ is the presence of an additional $\cosh(mh)$ term in the denominator of $p_b$.

The main differences between passive and active generation processes are that (i) the wave amplitudes and velocities
obtained with the instantly moving bottom are lower than
those generated by initial translation of the bottom motion and that (ii) the water column plays the role of
a low-pass filter (compare equations (\ref{active_eta})--(\ref{active_w}) with equations (\ref{genintsolbis})--(\ref{vervel})). 
High frequencies are attenuated in the moving bottom solution. Ward \cite{Ward}, who studied landslide
tsunamis, also commented on the $1/\cosh(mh)$ term, which low-pass filters the source spectrum. So the filter favors
long waves. In the discussion section, 
we will come back to the differences between passive generation and active generation. 

\subsection{Linear numerical method}

All the expressions derived from linear theory are explicit but they must be computed numerically. 
It is not a trivial task because of the oscillatory behaviour of the integrand functions.
All integrals were computed with Filon type numerical integration formulas \cite{Filon},
which explicitly take into account this oscillatory behaviour. Numerical results will be shown in Section 6.

\section{Nonlinear shallow water equations}

Synolakis and Bernard \cite{Syno2006} introduced a clear distinction between the various shallow-water models. At the lowest order of
approximation, one obtains the linear shallow water wave equation. The next level of approximation provides the 
nondispersive nonlinear shallow water equations (NSW). In the next level, dispersive terms are added and the
resulting equations constitute the Boussinesq equations. Since there are many different ways to go to this level
of approximation, there are a lot of different types of Boussinesq equations.  
The NSW equations are the most commonly used equations for tsunami propagation (see in particular the code MOST developed by the 
National Oceanic and Atmospheric Administration in the US \cite{TS} or the code TUNAMI developed by the 
Disaster Control Research Center in Japan). They are also used for generation and runup/inundation. For wave runup,
the effects of bottom friction become important and must be included in the codes. 
Our analysis will focus on the NSW equations. For simplicity, we assume below that $h$ is constant. Therefore
one can take $h$ as reference depth, so that the seafloor is given by $z=-1+\varepsilon\zeta$.

\subsection{Mathematical model}
In this subsection, partial derivatives are denoted by subscripts. 
When $\mu^2$ is a small parameter, the water is considered to be shallow. For the shallow water theory, 
one formally expands the potential $\phi$ in powers of $\mu^2$:
$$ \phi = \phi_0 + \mu^2 \phi_1 + \mu^4 \phi_2 + \cdots .$$
This expansion is substituted into the governing equation and the boundary conditions. The lowest-order
term in Laplace's equation is
\begin{equation}
\phi_{0zz} = 0.
\end{equation}
The boundary conditions imply that $\phi_0=\phi_0(x,y,t)$. Thus the vertical velocity component
is zero and the horizontal velocity components are independent of the vertical coordinate $z$
at lowest order. Let us introduce the notation $u:=\phi_{0x}(x,y,t)$ and $v:=\phi_{0y}(x,y,t)$. Solving Laplace's equation and taking
into account the bottom kinematic condition yield the following expressions for $\phi_1$ and $\phi_2$:
\begin{eqnarray*}
\phi_1(x,y,z,t)  &= & -\frac{1}{2} Z^2(u_x+v_y) + z \left[\zeta_t + \varepsilon (u\zeta_x + v\zeta_y)\right], \\
\phi_2(x,y,z,t) & = &\frac{1}{24} Z^4(\Delta u_x +\Delta v_y)
                     + \varepsilon \left(\varepsilon \frac{z^2}{2} |\nabla \zeta|^2 -
                \frac{1}{6} Z^3 \Delta\zeta\right)(u_x+v_y) \\
                   & & -\frac{\varepsilon}{3}Z^3 \nabla\zeta \cdot \nabla (u_x+v_y)
                   -\frac{z^3}{6}\bigl(\Delta\zeta_t+\varepsilon\Delta(u\zeta_x+v\zeta_y)\bigr)+\\
                && z(-1+\varepsilon\zeta)\left[\varepsilon\nabla\zeta \cdot \nabla\bigl(\zeta_t + \varepsilon (u\zeta_x +v\zeta_y)\bigr)-
                   \varepsilon^2|\nabla\zeta|^2(u_x+v_y) - \right. \\
&& \left.\frac{1}{2}(-1+\varepsilon\zeta)\bigl(\Delta\zeta_t + \varepsilon \Delta(u\zeta_x + v\zeta_y)\bigr)\right],
\end{eqnarray*}
where $$ Z=1+z-\varepsilon \zeta.$$

The next step consists in retaining terms of requested order in the free-surface boundary conditions. 
Powers of $\varepsilon$ will appear when expanding in Taylor series the free-surface conditions around
$z=0$. For example, if one keeps terms of order $\varepsilon\mu^2$ and $\mu^4$ in the dynamic boundary 
condition (\ref{dynfreesurfadim}) and in the kinematic boundary condition (\ref{kinfreesurfadim}), one obtains
\begin{equation}
\label{dyn1} 
\mu^2\phi_{0t} -  \frac{1}{2}\mu^4 (u_{tx}+v_{ty}) + \mu^2\eta + \frac1{2} \varepsilon \mu^2 (u^2+v^2)  =  0,
\end{equation}
\begin{multline}
 \mu^2 [\eta_t + \varepsilon(u\eta_x+v\eta_y) + \bigl(1+\varepsilon(\eta-\zeta)\bigr)(u_x+v_y)-\zeta_t-\varepsilon(u\zeta_x+v\zeta_y)] =\\
 \frac1{6} \mu^4(\Delta u_x + \Delta v_y). \label{kin1}
\end{multline}
Differentiating (\ref{dyn1}) first with respect to $x$ and then with respect to $y$ gives a set of two
equations:
\begin{eqnarray}
\label{dynx} u_t + \varepsilon (uu_x+vv_x) + \eta_x - \frac1{2} \mu^2 (u_{txx}+v_{txy}) & = & 0, \\
\label{dyny} v_t + \varepsilon (uu_y+vv_y) + \eta_y - \frac1{2} \mu^2 (u_{txy}+v_{tyy}) & = & 0.
\end{eqnarray}
The kinematic condition (\ref{kin1}) becomes
\begin{equation}
\label{kin2bis} (\eta-\zeta)_t + [u(1+\varepsilon(\eta-\zeta))]_x + [v(1+\varepsilon(\eta-\zeta))]_y = \frac1{6} \mu^2
(\Delta u_x + \Delta v_y).
\end{equation}
Equations (\ref{dynx}),(\ref{dyny}) and (\ref{kin2bis}) contain in fact various shallow-water models. The so-called
fundamental NSW equations which contain no dispersive effects are obtained by neglecting the terms of order $\mu^2$:
\begin{eqnarray}
\label{sw1} u_t + \varepsilon (uu_x+vu_y) + \eta_x & = & 0, \\
\label{sw2} v_t + \varepsilon (uv_x+vv_y) + \eta_y & = & 0, \\
\label{sw3} \eta_t + [u(1+\varepsilon(\eta-\zeta))]_x + [v(1+\varepsilon(\eta-\zeta))]_y & = & \zeta_t.
\end{eqnarray}
Going back to a bathymetry $h^*(x,y,t)$ equal to $1-\varepsilon\zeta(x,y,t)$ and using the fact that $(u,v)$ is the
horizontal gradient of $\phi_0$, one can rewrite the system of NSW equations as
\begin{eqnarray}
\label{cg1} u_t + \frac{\varepsilon}{2}(u^2+v^2)_x +\eta_x & = & 0, \\
\label{cg2} v_t +  \frac{\varepsilon}{2}(u^2+v^2)_y + \eta_y & = & 0, \\
\label{cg3} \eta_t + [u(h^*+\varepsilon\eta)]_x + [v(h^*+\varepsilon\eta)]_y & = & -\frac1{\varepsilon} h^*_t.
\end{eqnarray}
The system of equations (\ref{cg1})--(\ref{cg3}) has been used for example by Titov and Synolakis \cite{TS} for the numerical
computation of tidal wave run-up. Note that this model does not include any bottom
friction terms.

The NSW equations with dispersion (\ref{dynx})--(\ref{kin2bis}), also known as the Boussinesq equations, can be written 
in the following form: 
\begin{eqnarray}
\label{cg1bis} u_t + \frac{\varepsilon}{2}(u^2+v^2)_x +\eta_x - \frac1{2} \mu^2 \Delta u_t & = & 0, \label{bous1} \\
\label{cg2bis} v_t +  \frac{\varepsilon}{2}(u^2+v^2)_y + \eta_y - \frac1{2} \mu^2 \Delta v_t & = & 0, \label{bous2} \\
\label{cg3bis} \eta_t + [u(h^*+\varepsilon\eta)]_x + [v(h^*+\varepsilon\eta)]_y - \frac1{6} \mu^2(\Delta u_x + \Delta v_y) 
& = & - \frac1{\varepsilon} h^*_t. \label{bous3}
\end{eqnarray}
Kulikov et al. \cite{Kulikov} have argued that the satellite altimetry observations of the Indian Ocean tsunami show some dispersive 
effects. However the steepness is so small that the origin of these effects is questionable. Guesmia et al. \cite{Mariotti} compared 
Boussinesq and
shallow-water models and came to the conclusion that the effects of frequency dispersion are minor. As pointed out in \cite{Synolakis},
dispersive effects are necessary only when examining steep gravity waves, which are not encountered in the
context of tsunami hydrodynamics in deep water. However they can be encountered in experiments such as those
of Hammack \cite{Hammack} because the parameter $\mu$ is much bigger. 

\subsection{Numerical method}

In order to solve the NSW equations, a finite-volume approach is used. For example LeVeque \cite{LeVeque} used a high-order
finite volume scheme to solve a system of NSW equations. Here the flux scheme we use is the
characteristic flux scheme, which was introduced by Ghidaglia et al. \cite{Ghida}. This numerical method 
satisfies the conservative properties at the discrete level. The NSW equations (\ref{cg1})--(\ref{cg3}) can be rewritten in the 
following conservative form:
\begin{equation} \label{NSWEc}
\pd{\wv}{t} + \pd{\Fv(\wv)}{x} + \pd{\Gv(\wv)}{y} = \Sv(x,y,\wv,t), 
\end{equation}
where
\begin{eqnarray}
\wv & = & (\eta,u,v), \\
\Fv & = & \left(u(h^*+\varepsilon\eta), \frac{\varepsilon}{2}(u^2+v^2) +\eta, 0\right), \\
\Gv & = & \left(v(h^*+\varepsilon\eta), 0, \frac{\varepsilon}{2}(u^2+v^2) +\eta\right), \\
\Sv & = & \left(-\zeta_t, 0, 0\right).
\end{eqnarray}

The scheme we use is multi-dimensional by construction and does not require the solution of any Riemann problem.
For the sake of simplicity in the description of the numerical method, we assume that there is no $y$-variation and no source
term $\Sv$. Let us then consider a system of $m$--conservation laws ($m\geq1$)
\begin{equation}
\label{vffc1} \pd{\wv}{t} + \pd{\Fv(\wv)}{x} = 0, \qquad x \in \mathbb{R} ,\qquad t \geq 0,
\end{equation}
where $\wv \in \R^m$ and $\Fv: \R^m \mapsto \R^m$. We denote by $\Av(\wv)$ the Jacobian matrix of $\Fv(\wv)$:
\begin{equation}
\label{vffc10} \Av_{ij}(\wv) = \pd{\Fv_i}{\wv_j}(\wv), \qquad 1\leq i, j \leq m. 
\end{equation}
The system (\ref{vffc1}) is assumed to be hyperbolic. In other words, for every $\wv$ there exists a smooth basis
$(r_1(\wv),\dots,r_m(\wv))$ of $\R^m$ consisting of eigenvectors of $\Av(\wv)$. Said differently,
there exists $\lambda_k(\wv) \in \R$ such that $\Av(\wv)r_k(\wv) = \lambda_k(\wv)r_k(\wv)$. It is then
possible to construct $(\ell_1(\wv),\dots,\ell_m(\wv))$ such that 
$$ ^t\Av(\wv)\ell_k(\wv) = \lambda_k(\wv)\ell_k(\wv) \quad \mbox{and} \quad \ell_k(\wv) \cdot r_p(\wv) = \delta_{kp}. $$

Let $\R=\cup_{j\in\Z}[x_{j-1/2},x_{j+1/2}]$ be a one-dimensional mesh. Let also $\R_+=\cup_{n\in\N}[t_n,t_{n+1}]$. 
Let us discretize (\ref{vffc1}) by a finite volume method. We set 
$$ \Delta x_j = x_{j+1/2} - x_{j-1/2}, \quad  \Delta t_n = t_{n+1} - t_n $$
and
$$ \widetilde\wv_j^n = \frac{1}{\Delta x_j} \int_{x_{j-1/2}}^{x_{j+1/2}} \wv(x,t_n)\,dx, \quad
\widetilde\Fv_{j+1/2}^n = \frac{1}{\Delta t_n} \int_{t_n}^{t_{n+1}} \Fv\left(\wv(x_{j+1/2},t)\right)\,dt. $$
The system (\ref{vffc1}) can then be rewritten (exactly) as
\begin{equation}\label{vffc3} 
\widetilde\wv^{n+1}_{j} = \widetilde\wv^{n}_{j} - \frac{\Delta t_n}{\Delta x_j} (\widetilde\Fv^{n}_{j+1/2}-\widetilde\Fv^{n}_{j-1/2}).
\end{equation}
For a three-point explicit numerical scheme one has
\begin{equation}
\label{vffc5} \widetilde\Fv^{n}_{j+1/2} \approx \fv_j^n(\wv^{n}_{j},\wv^{n}_{j+1}),
\end{equation}
where the function $\fv$ is to be specified.
Multiplying (\ref{vffc1}) by $\Av(\wv)$ yields
\begin{equation}
\label{vffc11} \pd{\Fv(\wv)}{t} + \Av(\wv) \pd{\Fv(\wv)}{x} = 0.
\end{equation}
This shows that the flux $\Fv(\wv)$ is advected by $\Av(\wv)$. The numerical flux $\fv_j^n(\wv^{n}_{j},\wv^{n}_{j+1})$
represents the flux at an interface. Using a mean value $\mu^n_{j+1/2}$ of $\wv$ at this interface, we replace 
(\ref{vffc11}) by the linearization
\begin{equation}
\label{vffc10b} \pd{\Fv(\wv)}{t} + \Av(\mu^n_{j+1/2}) \pd{\Fv(\wv)}{x} = 0.
\end{equation}

We define the $k-$th characteristic flux component to be $F_k(\wv) = \ell_k(\mu^n_{j+1/2}) \cdot \Fv(\wv)$. It follows that
\begin{equation}
\label{vffc10c} \pd{F_k(\wv)}{t} + \lambda_k(\mu^n_{j+1/2}) \pd{F_k(\wv)}{x} = 0.
\end{equation}
This linear equation can be solved explicitly for $F_k(\wv)$. As a result it is natural to define the characteristic
flux $\fv^{CF}$ at the interface between two cells $[x_{j-1/2},x_{j+1/2}]$ and $[x_{j+1/2},x_{j+3/2}]$ as follows: for
$k \in {1,\dots,m}$, 
\begin{eqnarray*}
\ell_k(\mu^n_{j+1/2}) \cdot \fv_j^{CF,n}(\wv_j^n,\wv_{j+1}^n) & = & \ell_k(\mu^n_{j+1/2}) \cdot \Fv(\wv_j^n), \quad
\mbox{when} \;\; \lambda_k(\mu^n_{j+1/2}) > 0, \\
\ell_k(\mu^n_{j+1/2}) \cdot \fv_j^{CF,n}(\wv_j^n,\wv_{j+1}^n) & = & \ell_k(\mu^n_{j+1/2}) \cdot \Fv(\wv_{j+1}^n), \quad
\mbox{when} \;\; \lambda_k(\mu^n_{j+1/2}) < 0, \\
\ell_k(\mu^n_{j+1/2}) \cdot \fv_j^{CF,n}(\wv_j^n,\wv_{j+1}^n) & = & \ell_k(\mu^n_{j+1/2}) \cdot \left(\frac
{\Fv(\wv_j^n)+\Fv(\wv_{j+1}^n)}{2}\right), 
\end{eqnarray*}
when $\lambda_k(\mu^n_{j+1/2}) = 0$. Here
$$ \mu^n_{j+1/2} = \frac{\Delta x_j \wv_j^n + \Delta x_{j+1} \wv_{j+1}^n}{\Delta x_j + \Delta x_{j+1}}. $$

The characteristic flux can be written as
$$ \fv_j^{CF,n}(\wv_j^n,\wv_{j+1}^n) = \fv^{CF}(\mu_j^n;\wv_j^n,\wv_{j+1}^n) $$
where
\begin{equation} \label{vffc12} 
\fv^{CF}(\mu;\wv_1,\wv_2)= \frac{\Fv(\wv_1)+\Fv(\wv_2)}{2}-\mbox{sgn}\left(\Av(\mu(\wv_1,\wv_2)\right)\frac{\Fv(\wv_2)-\Fv(\wv_1)}{2}.
\end{equation}
The sign of the matrix $\Av(\mu)$ is defined by
$$ \mbox{sgn}(\Av(\mu))\Phi = \sum_{k=1}^{k=m} \mbox{sgn}(\lambda_k)(\ell_k(\mu) \cdot \Phi)r_k(\mu). $$

Going back to (\ref{vffc3}), one can construct the following explicit scheme:
\begin{equation}
\wv^{n+1}_{j} = \wv^{n}_{j} - \frac{\Delta t_n}{\Delta x_j} \left(\fv_j^{CF,n}(\wv_j^n,\wv_{j+1}^n)-\fv_j^{CF,n}(\wv_{j-1}^n,\wv_j^n)
\right).
\end{equation}

The characteristic flux scheme (\ref{vffc12}) gives very good results, especially when complex systems are considered 
\cite{Ghida}. In our case, we have to consider equation (\ref{vffc1}) in two dimensions and to discretise 
the source term too:
\begin{equation}
\pd{\wv}{t} + \pd{\Fv(\wv)}{x} + \pd{\Gv(\wv)}{y} = \Sv(x,y,\wv,t). 
\label{vffc13}
\end{equation}
One can refer to \cite{GhidaEJMB} for these two extensions. 

\section{Numerical method for the full equations}

The fully nonlinear potential flow (FNPF) equations (\ref{lapladim})--(\ref{dynfreesurfadim}) are solved by 
using a numerical model based on the Boundary Element Method (BEM). An accurate code
was developed by Grilli et al. \cite{Grilli}. It uses a
high-order three-dimensional boundary element method combined with
mixed Eulerian--Lagrangian time updating, based on second-order
explicit Taylor expansions with adaptive time steps. The efficiency of the code
was recently greatly improved by introducing a 
more efficient spatial solver, based on the fast multipole
algorithm \cite{Foche}. By replacing every matrix--vector product of the
iterative solver and avoiding the building of the influence
matrix, this algorithm reduces the computing complexity from
$O(N^2)$ to nearly $O(N)$ up to logarithms, where $N$ is the number of nodes on the boundary. 

By using Green's second identity, Laplace's equation (\ref{Laplace3}) is transformed into the boundary integral equation
\begin{equation}
\alpha(\textbf{x}_l)\phi(\textbf{x}_l) = \int_\Gamma \left(\pd{\phi}{n}(\textbf{x})G(\textbf{x},\textbf{x}_l) - 
\phi(\textbf{x})\pd{G}{n}(\textbf{x},\textbf{x}_l)\right)\,d\Gamma,
\end{equation}
where $G$ is the three-dimensional free space Green's function. The notation $\partial G/\partial n$ means the normal derivative,
that is $\partial G/\partial n=\nabla G \cdot {\bf n}$, with \textbf{n} the unit outward normal vector.
The vectors $\textbf{x} = (x,y,z)$ and $\textbf{x}_l = (x_l,y_l,z_l)$ 
are position vectors for points on the boundary, and 
$\alpha (\textbf{x}_l) = \theta_l / (4\pi)$ is a geometric coefficient, with $\theta_l$ the exterior solid angle made by 
the boundary at point $\textbf{x}_l$.
The boundary $\Gamma$ is divided into various parts with different boundary conditions. On the free surface, one rewrites
the nonlinear kinematic and dynamic boundary conditions in a mixed Eulerian-Lagrangian form,
\begin{eqnarray}
\label{surf1} \god{\textbf{R}}{t}&=&\nabla \phi, \\
\label{surf2} \god{\phi}{t}&=&-gz+\frac1{2}\nabla\phi \cdot \nabla\phi, 
\end{eqnarray}
 with \textbf{R} the position vector of a free-surface fluid 
particle. The material derivative is defined as 
\begin{equation}
\god{}{t}=\pd{}{t}+\textbf{q} \cdot \nabla.
\end{equation}

For time integration, second-order explicit Taylor series
expansions are used to find the new position and the potential on
the free surface at time $t+\Delta t$. 
This time stepping scheme presents the advantage of being explicit,
and the use of spatial derivatives along the free surface provides
a better stability of the computed solution.

The integral equations are solved by BEM. The boundary is
discretized into $N$ collocation nodes and $M$ high-order elements
are used to interpolate between these nodes. Within each
element, the boundary geometry and the field variables are
discretized using polynomial shape functions. The integrals on the boundary are converted into a sum on the
elements, each one being calculated on the reference element. The matrices are built with the numerical computation of the
integrals on the reference element. The linear systems resulting from the two boundary integral equations (one for the
pair $(\phi,\partial\phi/\partial n)$ and one for the pair $(\partial\phi/\partial t,\partial^2\phi/\partial t\partial n)$)
are full and non symmetric. Assembling the matrix as well as performing the
integrations accurately are time consuming tasks. They are done
only once at each time step, since the same matrix is used for
both systems. Solving the linear system is another time consuming
task. Even with the GMRES algorithm with preconditioning, the
computational complexity is $O(N^2)$, which is the same as the
complexity of the assembling phase. The introduction of the fast multipole algorithm reduces considerably
the complexity of the problem. The matrix
is no longer built. Far away nodes are placed in groups, so less time
is spent in numerical integrations and memory requirements are
reduced. The hierarchical structure involved in the
algorithm gives automatically the distance criteria for adaptive
integrations. 

Grilli et al. \cite{Vogel} used the earlier version of the code to study tsunami generation by underwater landslides. They included
the bottom motion due to the landslide. For the comparisons shown below, we only used the passive approach: we did
not include the dynamics of the bottom motion.

\section{Comparisons and discussion}

The passive generation approach is followed for the numerical comparisons between the three models: (i) linear equations, 
(ii) NSW equations and (iii) fully nonlinear equations.
As shown in Section 3, this generation process gives the largest transient-wave amplitudes for a given
permanent deformation of the seafloor. Therefore it is in some sense a worst case scenario. 

The small dimensionless numbers $\varepsilon$ and $\mu^2$ introduced in (\ref{sdnumbers}) represent the magnitude of  
the nonlinear terms and dispersive terms in the governing equations, respectively. Hence, the relative importance of the 
nonlinear and the dispersive effects is given by the parameter 
\begin{equation}
S = \frac{\mbox{nonlinear  terms}}{\mbox{dispersive  terms}} = \frac{\varepsilon}{\mu^2} = \frac{a\lambda^2}{d^3},
\end{equation}
which is called the Stokes (or Ursell) number \cite{Ursell}.\footnote{One finds sometimes in the literature a subtle difference
between the Stokes and Ursell numbers. Both involve a wave amplitude multiplied by the square of a wavelength divided
by the cube of a water depth. The Stokes number is defined specifically for the excitation of a closed basin
while the Ursell number is used in a more general context to describe the evolution of a long wave system. Therefore
only the characteristic length is different. For the Stokes number the length is the usual wavelength $\lambda$ related to the
frequency $\omega$ by $\lambda \approx 2\pi\sqrt{gd}/\omega$. In the Ursell number, the length refers
to the local wave shape independent of the exciting conditions.} An important assumption in the derivation of the Boussinesq
system (\ref{bous1})--(\ref{bous3}) is that the Ursell number is $O(1)$. Here, the symbol $O(\cdot)$ is used informally
in the way that is common in the construction and formal analysis of model equations for physical phenomena. We are 
concerned with the limits $\varepsilon \to 0$ and $\mu \to 0$. Thus, $S=O(1)$ means that, as $\varepsilon \to 0$ and $\mu \to 0$,
$S$ takes values that are neither very large nor very small. We emphasize here that the Ursell number does not convey any
information by itself about the separate negligibility of nonlinear and frequency dispersion effects. Another important
aspect of models is the time scale of their validity. In the NSW equations, terms of order $O(\varepsilon^2)$ and 
$O(\mu^2)$ have been neglected. Therefore one expects these terms to make an order-one relative contribution on a 
time scale of order min$(\varepsilon^{-2},\mu^{-2})$. 

\begin{figure}
\centerline{\includegraphics[width=1.2\linewidth]{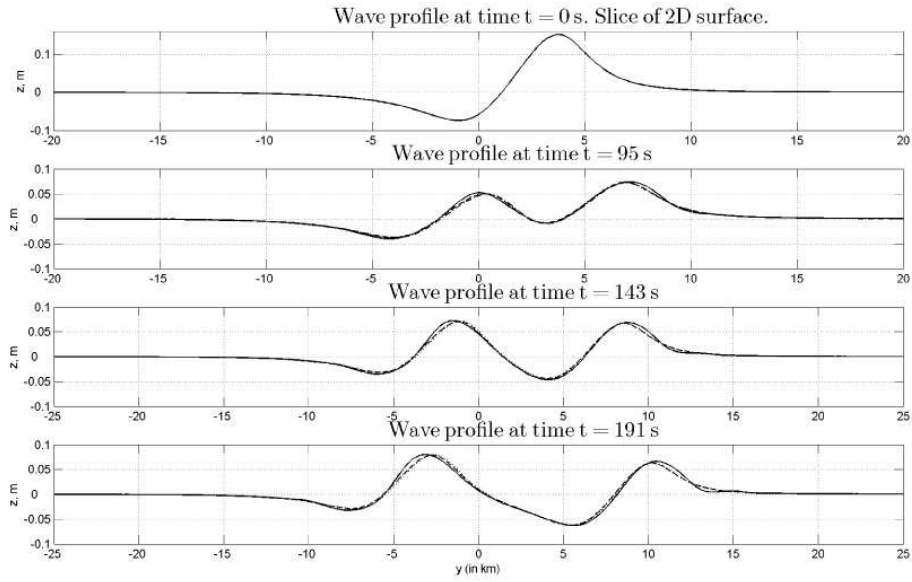}}
  \caption{Comparisons of the free-surface elevation at $x=0$ resulting from the integration of the linear equations ($\cdots$), 
NSW equations ($--$) and nonlinear equations ($-$) at different times of the propagation of transient waves generated by an earthquake
($t=0$ s, $t=95$ s, $t=143$ s, $t=191$ s). The parameters for the earthquake are those given in Table 1. The water depth is $h=100$ m.
One has the following estimates: $\varepsilon=5\times 10^{-4}$, $\mu^2=10^{-4}$ and consequently $S=5$.}
  \label{fig:ursell5}
\end{figure}

All the figures shown below are two-dimensional plots for convenience but we recall that all computations for the three 
models are three-dimensional.
Figure \ref{fig:ursell5} shows profiles of the free-surface elevation along the main direction of propagation ($y-$axis) of
transient waves generated by a permanent seafloor deformation corresponding to the parameters given in Table \ref{parset}. This 
deformation, which has been plotted in figure \ref{dip}, has been translated to the free surface. The water depth is 100 m. The 
small dimensionless numbers are roughly $\varepsilon=5\times 10^{-4}$ and $\mu=10^{-2}$, with a corresponding Ursell number equal 
to 5. One can see that the front system splits in two and propagates in both directions, with a leading wave of depression to the left 
and a leading wave of elevation to the right, in qualitative agreement with the satellite and tide gauge measurements of the 2004 
Sumatra event. When tsunamis are generated along subduction zones, they usually split in two; one moves quickly inland while the
second heads toward the open ocean. 
The three models are almost undistinguishable at all times: the waves propagate with the same 
speed and the same profile. Nonlinear effects and dispersive effects are clearly negligible during the first moments of 
transient waves generated by a moving bottom, at least for these particular choices of $\varepsilon$ and $\mu$.

\begin{figure}
\centerline{\includegraphics[width=1.1\linewidth]{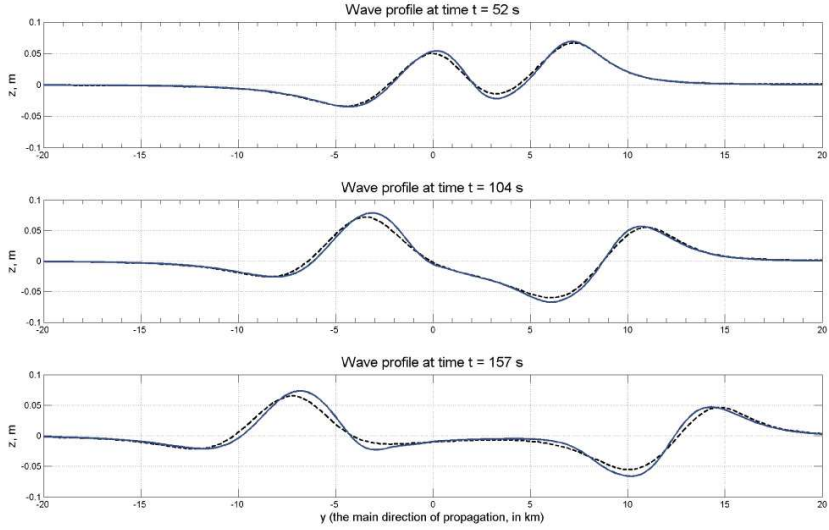}}
  \caption{Comparisons of the free-surface elevation at $x=0$ resulting from the integration of the linear equations ($\cdots$), 
NSW equations ($--$) and nonlinear equations ($-$) at different times of the propagation of transient waves generated by an earthquake
($t=52$ s, $t=104$ s, $t=157$ s). The parameters for the earthquake are those given in Table 1. The water depth is $h=500$ m.
One has the following estimates: $\varepsilon=10^{-4}$, $\mu^2=2.5 \times 10^{-3}$ and consequently $S=0.04$.}
  \label{fig:3times05}
\end{figure}

Let us now decrease the Ursell number by increasing the water depth. 
Figure \ref{fig:3times05} illustrates the evolution of transient water waves computed with 
the three models for the same parameters as those of figure \ref{fig:ursell5}, except for the water depth now equal to 500 m. The 
small dimensionless numbers are roughly $\varepsilon=10^{-4}$ and $\mu=5 \times 10^{-2}$, with a corresponding Ursell number 
equal to 0.04. The linear and nonlinear profiles cannot be distinguished within graphical accuracy. Only the NSW profile is
slightly different.

\begin{figure}
\centerline{\includegraphics[width=1.4\linewidth]{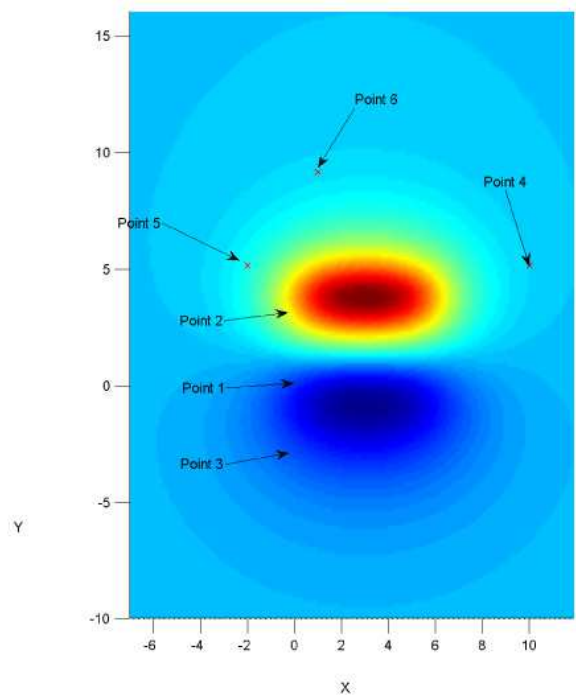}}
  \caption{Top view of the initial free surface deformation showing the location of six selected gauges, with the following
coordinates (in km): (1) 0,0 ; (2) 0,3 ; (3) 0,$-3$ ; (4) 10,5; (5) $-2$,5 ; (6) 1,10. The lower oval area 
represents the initial subsidence while the upper oval area represents the initial uplift.}
  \label{fig:capteurs}
\end{figure}

\begin{figure}
\centerline{\includegraphics[width=1.3\linewidth]{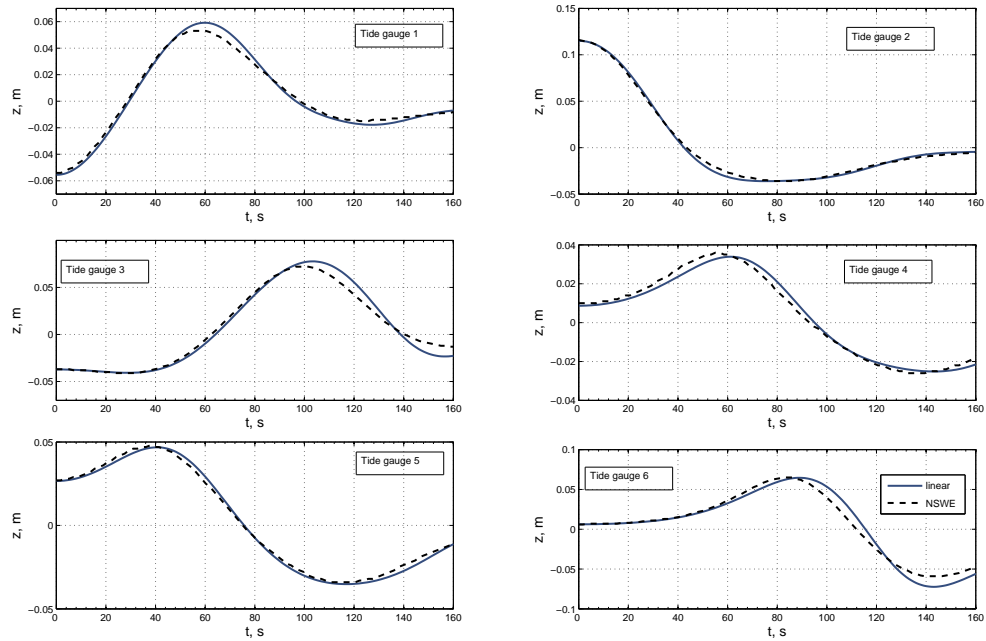}}
  \caption{Transient waves generated by an underwater earthquake. Comparisons of the free-surface elevation as a function of time 
at the selected gauges shown in figure \ref{fig:capteurs}: ${-}$, linear model ; $--$ nonlinear shallow water model. 
The time $t$ is expressed in seconds. The physical parameters are those of figure \ref{fig:3times05}. Since the fully nonlinear
results cannot be distinguished from the linear ones, they are not shown.}
  \label{fig:gages05}
\end{figure}

Let us introduce several sensors (tide gauges) at selected locations which are
representative of the initial deformation of the free surface (see figure \ref{fig:capteurs}).
One can study the evolution of the surface elevation during the generation time at each gauge. Figure
\ref{fig:gages05} shows free-surface elevations corresponding to the linear and nonlinear shallow water models. They are plotted 
on the same graph for comparison purposes. Again there is a slight difference between the linear and the NSW models,
but dispersion effects are still small.

\begin{figure}
\centerline{\includegraphics[angle=90,width=0.8\linewidth]{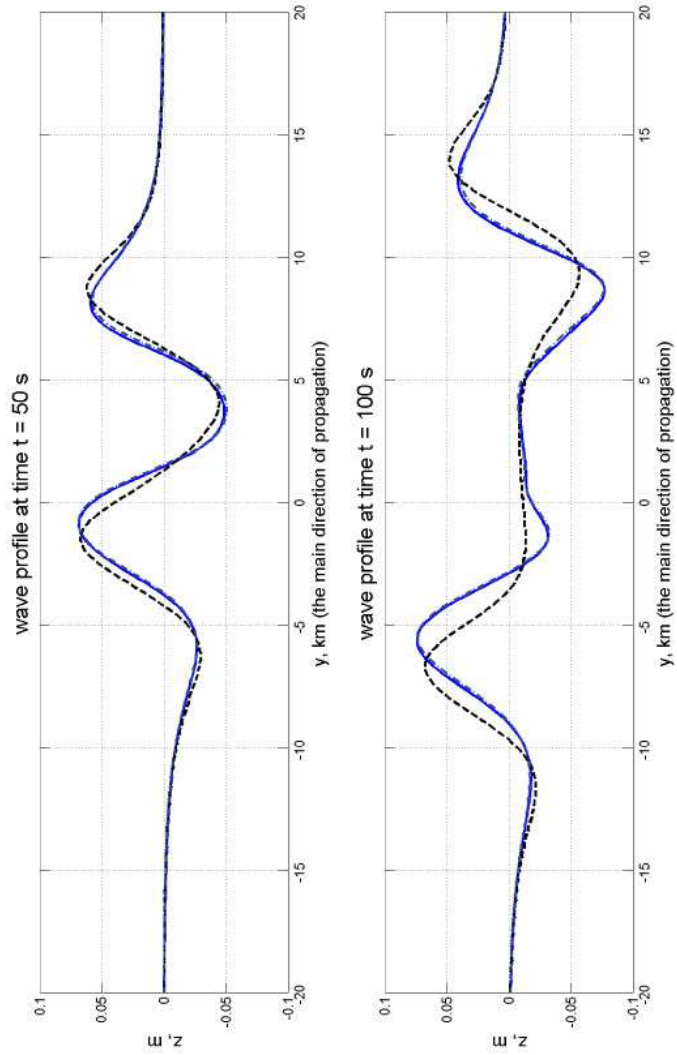}}
  \caption{Comparisons of the free-surface elevation at $x=0$ resulting from the integration of the linear equations ($-\cdot-$), 
NSW equations ($--$) and FNPF equations
(${-}$) at different times of the propagation of transient waves generated by an earthquake
($t=50$ s, $t=100$ s). The parameters for the earthquake are those given in Table 1. The water depth is 1 km.
One has the following estimates: $\varepsilon=5\times 10^{-5}$, $\mu^2=10^{-2}$ and consequently $S=0.005$.}
  \label{fig:2times1}
\end{figure}

\begin{figure}
\centerline{\includegraphics[angle=90,width=0.8\linewidth]{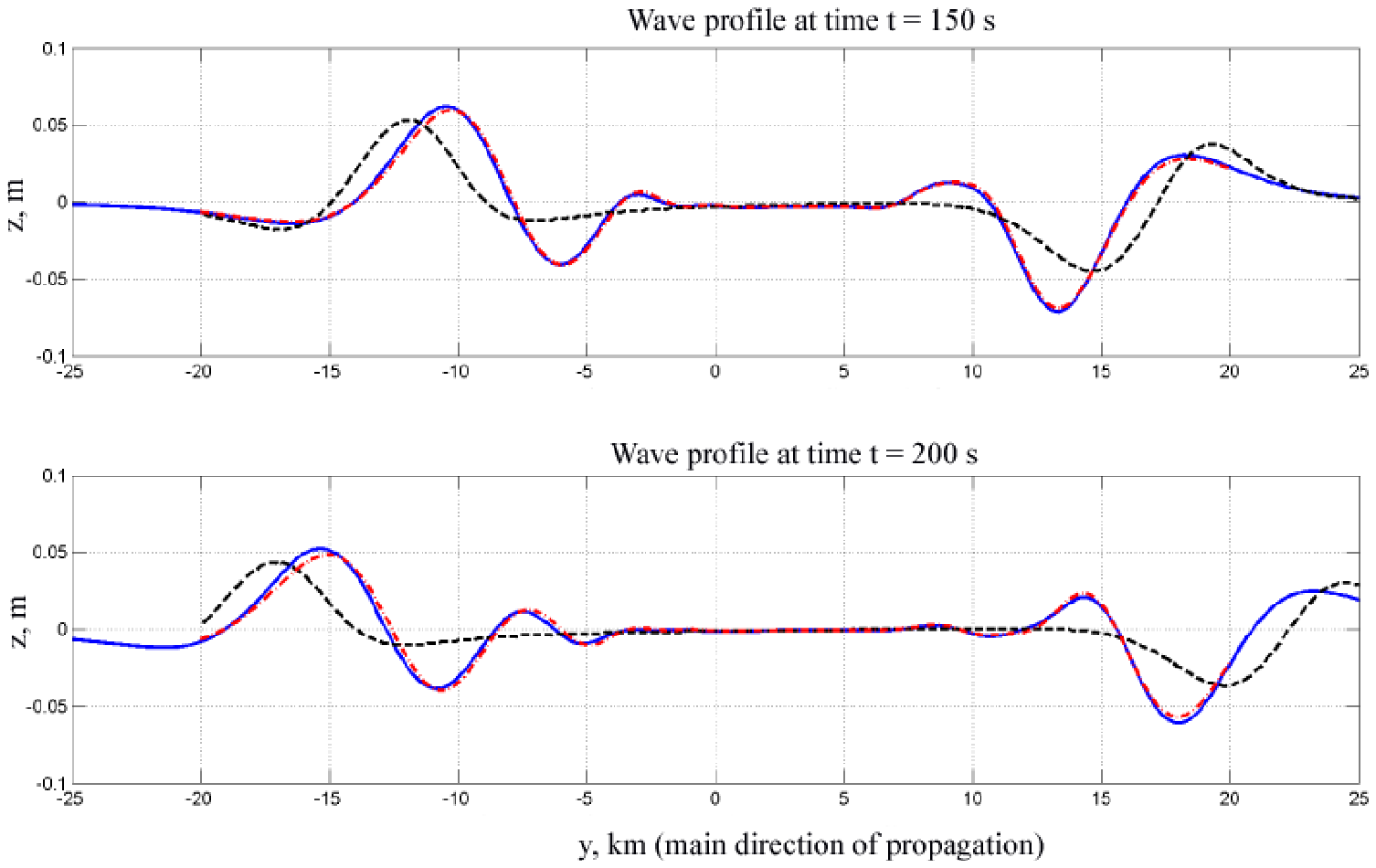}}
  \caption{Same as figure \ref{fig:2times1} for later times ($t=150$ s, $t=200$ s).}
  \label{fig:2times2}
\end{figure}

Let us decrease the Ursell number even further by increasing the water depth. Figures \ref{fig:2times1} and \ref{fig:2times2}
illustrate the evolution of transient water waves computed with 
the three models for the same parameters as those of Figure \ref{fig:ursell5}, except for the water depth now equal to 1 km. The 
small dimensionless numbers are roughly $\varepsilon=5 \times 10^{-5}$ and $\mu=0.1$, with a corresponding Ursell number 
equal to 0.005. On one hand, linear and fully nonlinear models are essentially undistinguishable at all times: the waves propagate
with the same speed and the same profile. Nonlinear effects are clearly negligible during the first moments of transient waves 
generated by a moving bottom, at least in this context. On the other hand, the numerical solution obtained with the NSW model 
gives slightly different results. Waves computed with this model do not 
propagate with the same speed and have different amplitudes compared to those obtained with the linear and 
fully nonlinear models. Dispersive effects come into the picture essentially because the waves are shorter compared to the
water depth. As shown in the previous examples, dispersive effects do not play a role for long enough waves.  

\begin{figure}
\centerline{\includegraphics[width=1.3\linewidth]{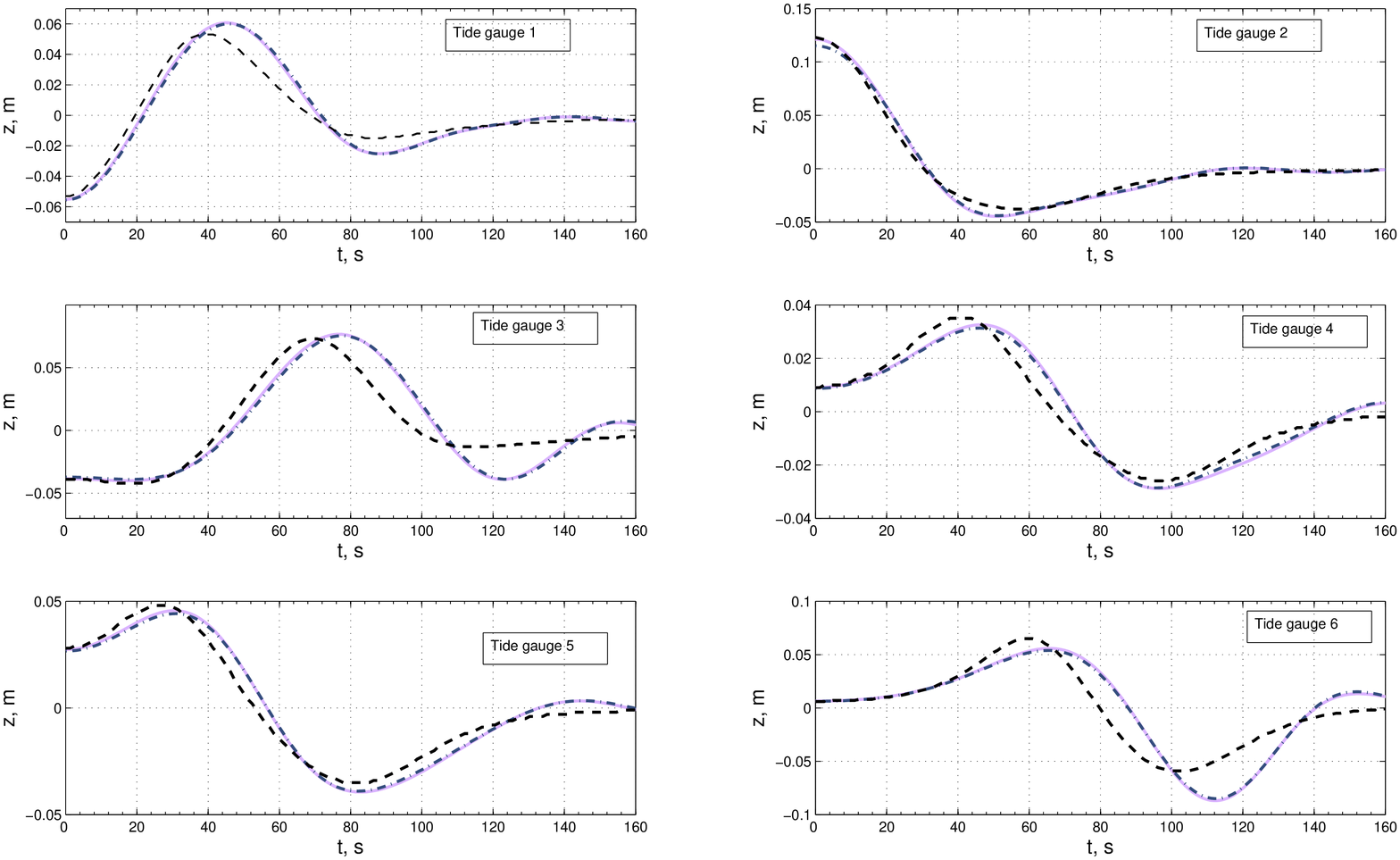}}
  \caption{Transient waves generated by an underwater earthquake. The physical parameters are those of figures \ref{fig:2times1} and
\ref{fig:2times2}. Comparisons of the free-surface elevation as a function of time 
at the selected gauges shown in figure \ref{fig:capteurs}: ${-}$, linear model ; $--$ nonlinear shallow water model. 
The time $t$ is expressed in seconds. The FNPF results cannot be distinguished from the linear results.}
\label{fig:gages}
\end{figure}

Figure \ref{fig:gages} shows the transient waves at the gauges selected in figure \ref{fig:capteurs}. 
One can see that the elevations obtained with the linear and fully nonlinear models are very close within graphical accuracy.
On the contrary, the nonlinear shallow water model leads to a higher speed and the difference is obvious for the points away
from the generation zone.

These results show that one cannot neglect the dispersive effects any longer. 
The NSW equations, which contain no dispersive effects, lead to different speed and amplitudes. 
Moreover, the oscillatory behaviour just behind the two front waves is no longer present. This oscillatory behaviour has been 
observed for the water waves computed with the linear and fully nonlinear models and is due to the presence of 
frequency dispersion.
So, one should replace the NSW equations with Boussinesq models which combine the two fundamentals effects of nonlinearity
and dispersion. Wei et al. \cite{Wei} provided comparisons for two-dimensional waves resulting from the integration of a  
Boussinesq model and the two-dimensional version of the FNPF model described above. 
In fact they used a fully nonlinear variant of the Boussinesq model, which predicts wave heights, phase speeds and particle 
kinematics more accurately than the standard weakly nonlinear approximation first derived by Peregrine \cite{Pere} and improved 
by Nwogu's modified Boussinesq model \cite{Nwogu}. We refer to the review \cite{Kirby} on Boussinesq models and their 
applications for a complete description of modern Boussinesq theory. 

From a physical point of view, we emphasize that the wavelength of the tsunami waves is directly related to the mechanism of 
generation and to the dimensions of the source event. And so is the dimensionless number $\mu$ which determines the importance 
of the dispersive effects. In general it will remain small. 

Adapting the discussion by Bona et al. \cite{BCL}, one can expect the solutions to the long wave models to be good approximations of the solutions
to the full water-wave equations on a time scale of the order min$(\varepsilon^{-1},\mu^{-2})$ and also the neglected effects to make
an order-one relative contribution on a time scale of order min$(\varepsilon^{-2},\mu^{-4},\varepsilon^{-1}\mu^{-2})$. Even though we 
have not computed precisely the constant in front of these estimates, the results shown in this paper are in agreement
with these estimates. Considering the 2004 Boxing Day tsunami, it is clear that dispersive and nonlinear effects did not have sufficient 
time to develop during the first hours due to the extreme smallness of $\varepsilon$ and $\mu^2$, except of course when the tsunami 
waves approached the coast.

Let us conclude this section with a discussion on the generation methods, which extends the results given in \cite{ddk}
\footnote{In figures 1 and 2 of \cite{ddk}, a mistake was introduced in the time scale. All times must be multiplied by a
factor $\sqrt{1000}$.}. We show the major differences between the classical passive approach and the active approach of wave generation 
by a moving bottom. Recall that the classical approach consists in translating the sea bed deformation to the free surface 
and letting it propagate. Results are presented for waves computed with the linear model.

\begin{figure}
	\centerline{\includegraphics[width=1.3\textwidth]{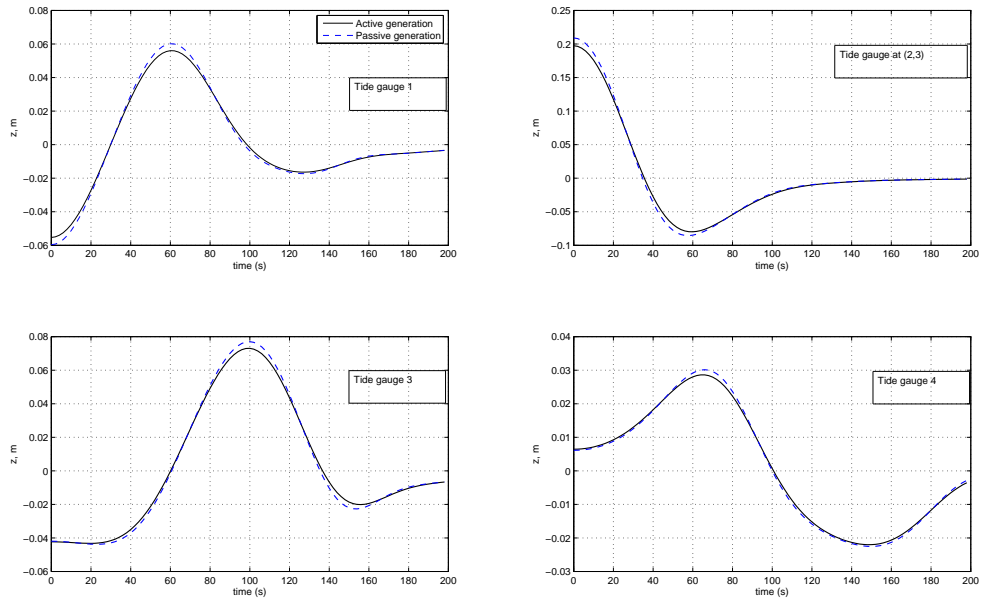}}
	\caption{Transient waves generated by an underwater earthquake. The computations are based on linear wave theory.
Comparisons of the free-surface elevation as a function of time 
at selected gauges for active and passive generation processes. The time $t$ is expressed in seconds. The physical parameters are those 
of figure \ref{fig:3times05}. In particular, the water depth is $h=500$ m.}
	\label{fig:gauge05}
\end{figure}

\begin{figure}
	\centerline{\includegraphics[width=1.2\textwidth]{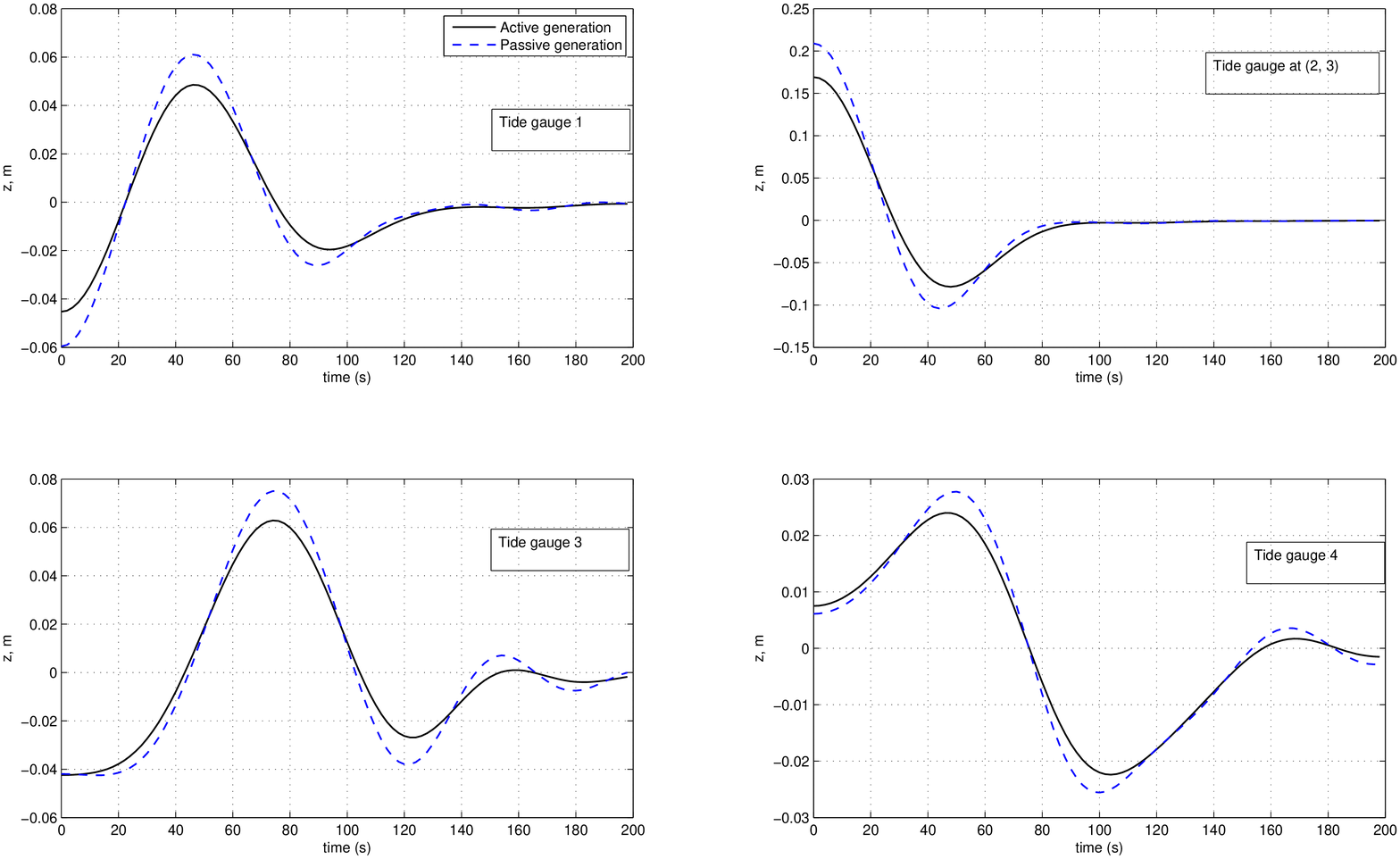}}
	\caption{Same as figure \ref{fig:gauge05}, except for the water depth, which is equal to 1 km.}
	\label{fig:gauge}
\end{figure}

Figure \ref{fig:gauge05} shows the waves measured at several artificial gauges. The parameters are those of Table \ref{parset},
and the water depth is $h=500$ m. The solid line represents 
the solution with an instantaneous bottom deformation while the dashed line represents the passive wave generation scenario. 
Both scenarios give roughly the same wave profiles. Let us now consider a slightly different set of parameters: the only
difference is the water depth which is now $h=1$ km. As shown in figure \ref{fig:gauge}, the 
two generation models differ. The passive mechanism gives higher wave amplitudes. 

Let us quantify this difference by considering the relative difference between the two mechanisms defined by
$$
  r(x,y,t) = \frac{\abs{\eta_{\rm active}(x,y,t)-\eta_{\rm passive}(x,y,t)}}{||{\eta_{\rm active}}||_\infty}.
$$
Intuitively this quantity represents the deviation of the passive solution from the active one with a 
moving bottom in units of the maximum amplitude of $\eta_{\rm active}(x,y,t)$.

\begin{figure}
	\centerline{\includegraphics[width=1.2\textwidth]{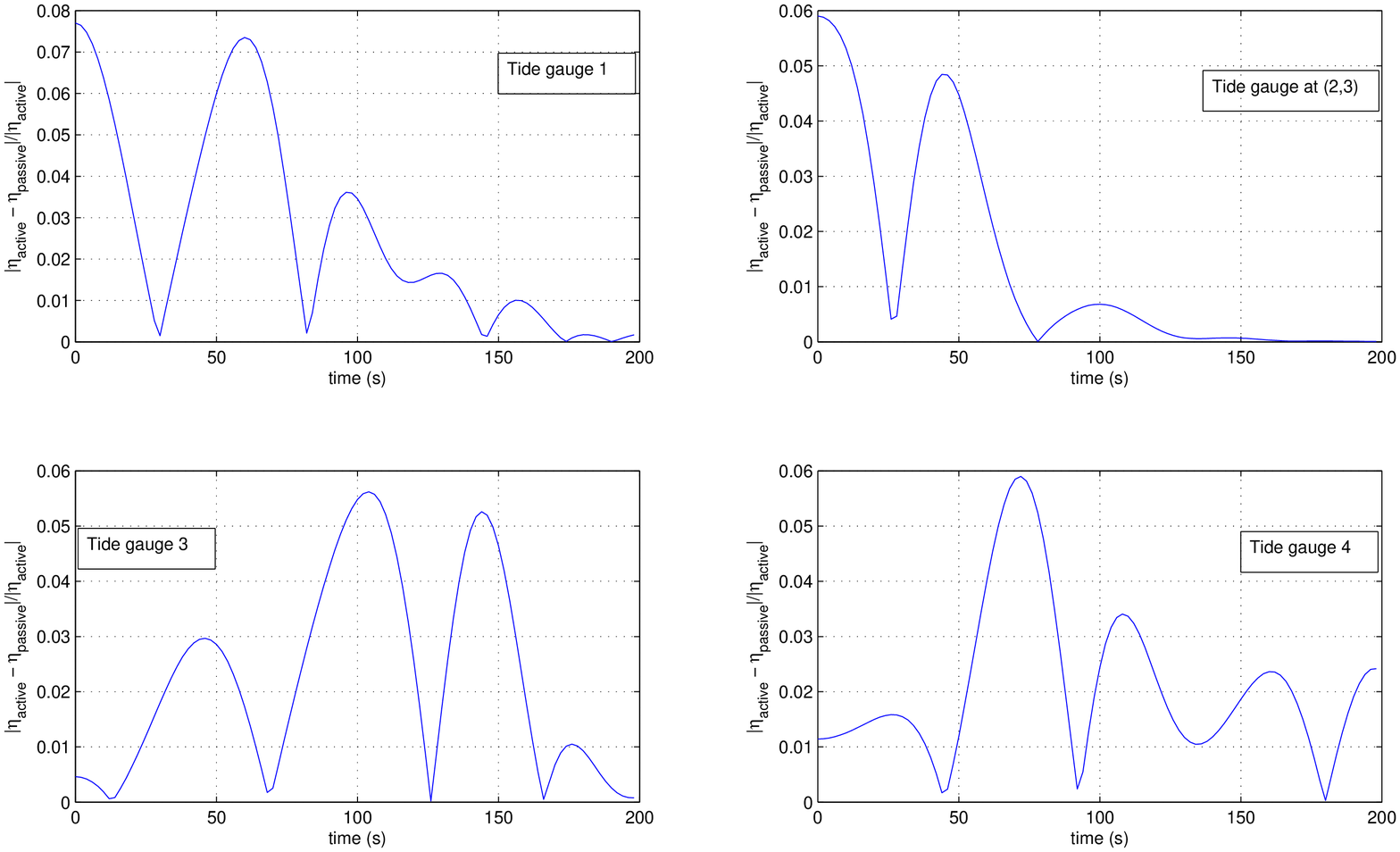}}
	\caption{Relative difference between the two solutions shown in figure \ref{fig:gauge05}. The time $t$ is expressed in seconds.}
	\label{fig:differ05}
\end{figure}

\begin{figure}
	\centerline{\includegraphics[width=1.2\textwidth]{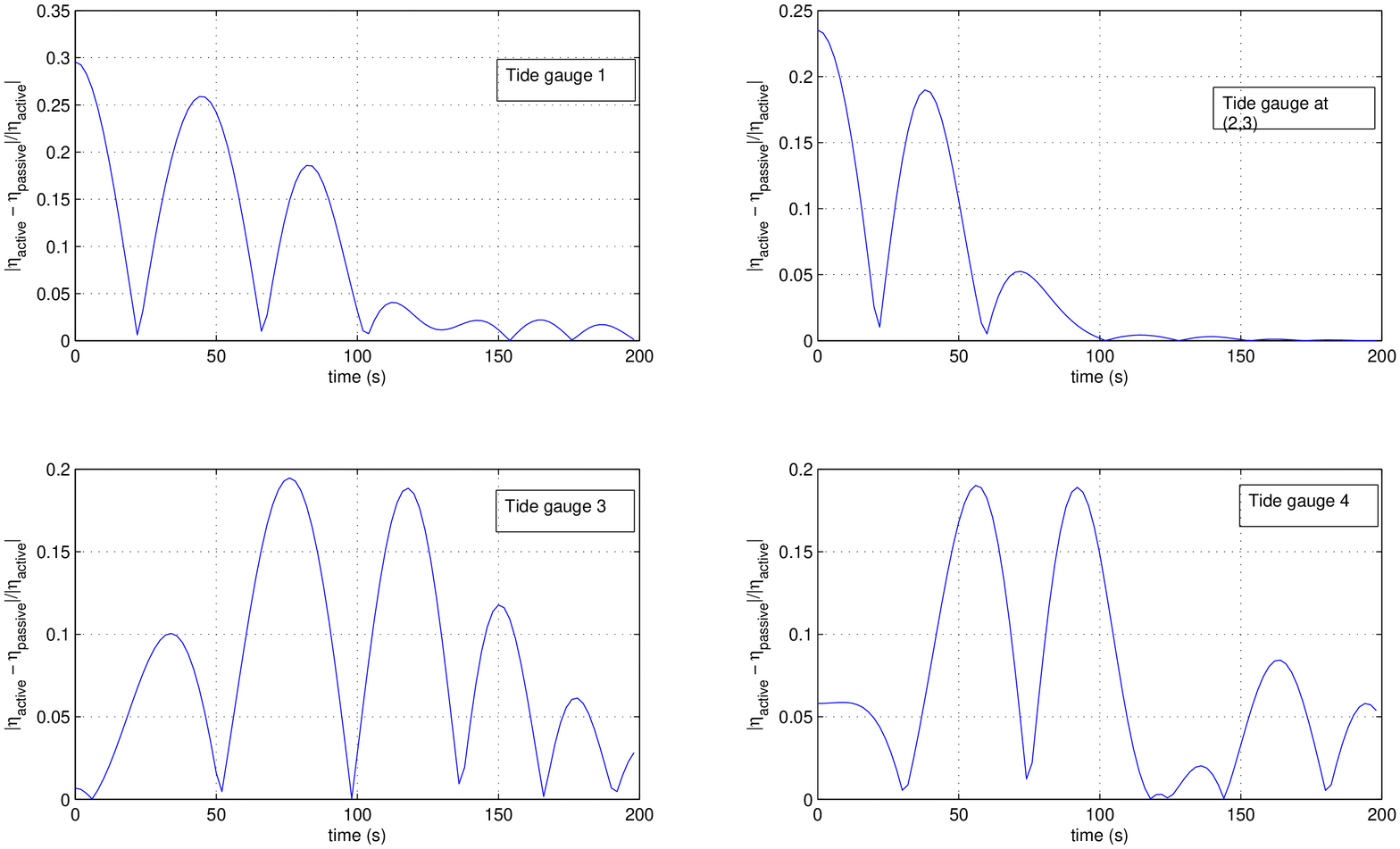}}
	\caption{Relative difference between the two solutions shown in figure \ref{fig:gauge}.}
	\label{fig:differ}
\end{figure}

Results are presented on figures (\ref{fig:differ05}) and (\ref{fig:differ}). The differences can be easily
explained by looking at the analytical formulas (\ref{active_eta}) and (\ref{genintsolbis}) of Section 3. These differences, which 
can be crucial for accurate tsunami modelling, are twofold.

First of all, the wave amplitudes obtained with the instantly moving bottom are lower than those generated by the passive
approach (this statement follows from the inequality $\cosh mh \geq 1$). The numerical experiments show that this difference 
is about $6\%$ in the first case and $20\%$ in the second case.

The second feature is more subtle. The water column has an effect of a low-pass filter. 
In other words, if the initial deformation contains high frequencies, they will be attenuated 
in the moving bottom solution because of the presence of the hyperbolic cosine $\cosh(mh)$ in the denominator which grows 
exponentially with $m$. Incidently, in the framework of the NSW equations, there is no difference between the passive and the 
active approach for an instantaneous seabed deformation \cite{Ernie,Tuck}. 

If we prescribe a more realistic bottom motion as in \cite{Dias2} for example, the results will 
depend on the characteristic time of the seabed deformation. When the characteristic time of the bottom motion decreases, the linearized 
solution tends to the instantaneous wave generation scenario. So, in the framework of linear water wave equations, one cannot
exceed the passive generation amplitude with an active process. However, during slow events, Todorovska and Trifunac \cite{todo} have 
shown that amplification of one order of magnitude may occur when the sea floor uplift spreads with velocity similar to the long 
wave tsunami velocity. 

\section {Conclusions}

Comparisons between linear and nonlinear models for tsunami generation by an underwater earthquake have been presented.
There are two main conclusions that are of great importance for modelling the first instants of a tsunami and for providing 
an efficient initial condition to propagation models. To begin with, a very good agreement is observed from the superposition of
plots of wave profiles computed with the linear and fully nonlinear models. Secondly, the nonlinear shallow water model was not 
sufficient to model some of the waves generated by a moving bottom because of the presence of frequency dispersion.
However classical tsunami waves are much longer, compared to the water depth, than the waves considered in the present paper,
so that the NSW model is also sufficient to describe tsunami generation by a moving bottom. 
Comparisons between the NSW equations and the FNPF equations for modeling tsunami run-up are left for future work. Another
aspect which deserves attention is the consideration of Earth rotation and the derivation of Boussinesq models in spherical
coordinates.





\section*{Acknowledgments}
The authors thank C. Fochesato for his help on the numerical method used to solve the fully nonlinear model. 
The first author gratefully acknowledges the kind assistance of the Centre de Math\'ematiques et de Leurs Applications 
of \'{E}cole Normale Sup\'erieure de Cachan. The third author acknowledges the support from the EU project TRANSFER 
(Tsunami Risk ANd Strategies For the European Region) of the 
sixth Framework Programme under contract no. 037058.

\end {document}